\documentclass[11pt]{article}
\usepackage{aas_macros,amsmath,amssymb,comment,cite,esint,graphicx,mathtools,diagbox}
\usepackage{bm}
\usepackage[margin=.8in,letterpaper]{geometry}
\usepackage[colorlinks=true]{hyperref}
\usepackage[affil-it]{authblk}
\usepackage[utf8]{inputenc}
\usepackage{subfigure}

\usepackage{mathrsfs}
\usepackage{appendix}
\usepackage{amssymb}
\usepackage{float}                  
\usepackage{color}
\usepackage{cite}
\usepackage{hyperref}
\hypersetup{pageanchor=false}
\usepackage{indentfirst}
\usepackage{url}
\usepackage{xfrac}
\usepackage{caption}
\usepackage[numbers,square,comma,sort&compress,merge]{natbib}
\usepackage{esint}
\usepackage{overpic}
\usepackage{graphicx}
\usepackage{epsf,amsmath,bbold,amsfonts,stmaryrd}
\usepackage{textcomp}
\usepackage{ulem}
\usepackage{tikz}
\usepackage{multirow}
\numberwithin{equation}{section}
\let\originalleft\left
\let\originalright\right
\renewcommand{\left}{\mathopen{}\mathclose\bgroup\originalleft}
\renewcommand{\right}{\aftergroup\egroup\originalright}

\def\bea{\begin{eqnarray}}
\def\eea{\end{eqnarray}}

\usepackage{tensor}
\usepackage{physics}

\newcolumntype{P}[1]{>{\Centering\hspace{0pt}}p{#1}}
\newcolumntype{Z}{>{\centering\arraybackslash}X} 

\newcommand{\df}{\mathrm{d}}   
   
\begin{document}
\title{\bf Imaging and Polarimetric Signatures of Konoplya–Zhidenko Black Holes with Various Thick Disk }
	
\author{Xinyu Wang$^{1}$, Yukang Wang$^{2}$, Xiao-Xiong Zeng$^{3\ast}$}
\date{}
	
\maketitle
\vspace{-15mm}

\begin{center}
{\it
$^1$ School of Physics and Astronomy, Beijing Normal University,
Beijing 100875, P. R. China\\\vspace{2mm}

$^2$ School of Material Science and Engineering, Chongqing Jiaotong University, Chongqing 400074, P. R. China\\\vspace{2mm}

$^3$ College of Physics and Electronic Engineering,
Chongqing Normal University, Chongqing 401331,  P. R. China \\\vspace{2mm}
}
\end{center}

\vspace{8mm}

\begin{abstract}

We investigate the imaging properties of spherically symmetric Konoplya–Zhidenko (KZ) black holes surrounded by geometrically thick accretion flows, adopting a phenomenological radiatively inefficient accretion flow (RIAF) model and an analytical ballistic approximation accretion flow (BAAF) model. General relativistic radiative transfer is employed to compute synchrotron emission from thermal electrons and generate horizon-scale images. For the RIAF model, we analyze the dependence of image morphology on the deformation parameter, observing frequency, and flow dynamics. The photon ring and central dark region expand with increasing deformation parameter, with brightness asymmetries arising at high inclinations and depending on flow dynamics and emission anisotropy. The BAAF disk produces narrower rings and darker centers, while polarization patterns trace the brightness distribution and vary with viewing angle and deformation, revealing spacetime structure. These results demonstrate that intensity and polarization in thick-disk models provide probes of KZ black holes and near-horizon accretion physics.

\end{abstract}

\vfill{\footnotesize $\ast$ Corresponding author: xxzengphysics@163.com}

\maketitle

\newpage
\baselineskip 18pt
\section{Introduction}\label{sec1}

Accumulating astronomical evidence strongly supports the existence of supermassive black holes (SMBHs) at galactic centers.  
Their presence has been confirmed through multiple independent observations, from the detection of gravitational waves by LIGO/Virgo~\cite{LIGOScientific:2016aoc} to the direct horizon-scale images captured by the Event Horizon Telescope (EHT)~\cite{EventHorizonTelescope:2019dse,EventHorizonTelescope:2022wkp}. 
Recent EHT polarization data~\cite{EventHorizonTelescope:2021bee,EventHorizonTelescope:2024hpu} further reveal detailed information about the magnetized plasma and radiation near event horizons, providing valuable probes of accretion physics and spacetime geometry.
Motivated by these developments, many studies have focused on modeling black hole accretion flow images within both general relativity (GR)~\cite{Hou:2023bep,Zhang:2024lsf,Zhang:2025vyx} and various modified gravity theories~\cite{Wang:2025buh,Chen:2025ysv,Li:2025knj,Hou:2022eev,Ahmed:2025ttq,NooriGashti:2024gnc,Guo:2023grt,Wan:2025gbm,Aslam:2025hgl,Zeng:2025kyv,Yang:2025whw,Ahmed:2025did,Ahmed:2025vww,Moriyama:2025isl}.

Einstein’s general relativity remains one of the most elegant and experimentally verified theories of nature~\cite{Will:2014kxa}. 
Nonetheless, current observations do not completely exclude the possibility of deviations from GR, leaving room for alternative theories. 
Konoplya and Zhidenko~\cite{Konoplya:2016pmh} proposed a rotating non-Kerr black hole metric incorporating a static deformation, which can be viewed as an axisymmetric vacuum solution of an as-yet-unknown modified gravity theory~\cite{Konoplya:2016jvv}. 
The motivation for this metric lies in exploring whether gravitational-wave observations might hint at departures from GR~\cite{Konoplya:2016pmh,Shashank:2021giy,Sakalli:2022xrb}. Several studies have shown that the Konoplya–Zhidenko (KZ) spacetime may describe a realistic astrophysical black hole~\cite{Bambi:2016iip,Ni:2016uik}, and that such non-Kerr geometries could modify strong-field signatures and observational phenomena. 
Recent works have examined particle dynamics \cite{Razzaq:2023fdd}, gravitational lensing \cite{Wang:2016paq,Nampalliwar:2019iti,Sakalli:2025dif}, and black hole shadows in the KZ background~\cite{Wang:2017hjl,He:2023hsv}, as well as the superradiant behavior of scalar and electromagnetic fields~\cite{Franzin:2021kvj}. 
Further investigations include tests against the hot-spot data of Sgr~A*~\cite{Shahzadi:2022rzq}, analyses of plasma effects~\cite{He:2023hsv} and relativistic viscous accretion flows~\cite{Patra:2024srh}, and studies of energy extraction via magnetic reconnection~\cite{Long:2024tws,Zhang:2024rvk}. 
More recently, the observational appearance and imaging features of the KZ black hole surrounded by a thin accretion disk have also been explored~\cite{He:2025rjq}.

SMBHs are generally believed to accrete hot, magnetized plasma, forming a luminous accretion disk whose thermal synchrotron emission likely dominates the observed black hole image frequencies \cite{Lu:2023bbn,EventHorizonTelescope:2019pgp}. The observed disk features reflect both gravitational effects and the underlying plasma properties, including the electron distribution, temperature, and magnetic field geometry~\cite{EventHorizonTelescope:2019pgp}. Previous imaging studies in KZ spacetimes have mainly focused on geometric lensing effects, often assuming simple, geometrically thin emission models~\cite{Luminet:1979nyg,He:2025rjq,Yang:2025usj,Zeng:2025pch,Zeng:2021dlj,Chael:2021rjo,Guerrero:2021ues,Hou:2022eev,Wang:2022yvi,Zhang:2023okw,Meng:2023htc,Gao:2023mjb}. 
However, observations by the EHT and other studies~\cite{EventHorizonTelescope:2019dse,EventHorizonTelescope:2019pgp,EventHorizonTelescope:2022urf,Ho:1999ss} have suggested that, under the dominant influence of gravity, accretion flows near SMBHs may become geometrically thick and optically thin, owing to the suppression of vertical cooling and compression of matter~\cite{Narayan:1994xi}. In such situations, it becomes essential to consider the detailed distributions of streamlines, particle number density, electron temperature, and magnetic field configuration.

Theoretical studies of geometrically thick accretion disks remain an active area of research~\cite{Abramowicz:2011xu}. 
Many works have focused on \textit{radiatively inefficient accretion flow} (RIAF) models~\cite{Yuan:2003dc,Broderick:2005at,Broderick:2008qf,Broderick:2010kx,Pu:2016qak,Pu:2018ute,Jiang:2023img,Saurabh:2025kwb,Yin:2025rao}, 
in which the vertically averaged electron density and temperature typically follow nearly power-law profiles with radius~\cite{Yuan:2003dc}. 
This phenomenological framework, physically motivated and broadly consistent with general relativistic magnetohydrodynamic (GRMHD) simulations, has been successful in reproducing the submillimeter spectra and overall image morphology of sources such as M87*~\cite{EventHorizonTelescope:2019pgp}. However, RIAF models neglect certain physical effects—such as outflows, non-thermal particles, and full GRMHD dynamics—which limit their predictive power, especially for polarization and variability studies~\cite{Yuan:2014gma,Broderick:2008sp,2009ApJ...706..497M}.

Traditionally, RIAF models have been employed to describe the large-scale behavior of accretion flows. 
However, the advent of horizon-resolving observations has enabled direct probes of plasma dynamics at event-horizon scales, making the study of near-horizon magnetofluids increasingly essential for understanding black hole environments. 
Recently, Hou \textit{et al.}~\cite{Hou:2023bep,Zhang:2024lsf} developed a self-consistent analytical model for horizon-scale accretion flows within the GRMHD framework, which we refer to as the \textit{ballistic approximation accretion flow} (BAAF) model.
Assuming that gravity dominates the fluid acceleration close to the event horizon, this model provides explicit expressions for thermodynamic variables and magnetic field configurations, offering a physically motivated description of the morphology and dynamics of geometrically thick accretion flows in the near-horizon region of black holes.

In this work, we investigate the imaging properties of the non-rotating KZ black hole using two representative accretion models: the phenomenological RIAF and the analytical BAAF. The images are computed using a general relativistic radiative transfer (GRRT) method. We analyze not only the intensity morphology under different parameters, viewing angles, and frequencies, but also the impact of emission anisotropy and distinct flow dynamics. Furthermore, we explore the polarization signatures predicted by the BAAF model.

The remaining sections of this paper are structured as follows.
In Sec.~\ref{sec2}, we briefly review the properties of non-rotating black holes in KZ gravity and derive the corresponding photon sphere.
Sec.~\ref{sec3} introduces the synchrotron radiation mechanism and the GRRT methodology.
In Sec.~\ref{sec4}, we analyze the imaging characteristics under the RIAF model, including the effects of emission anisotropy and distinct flow dynamics.
Sec.~\ref{sec5} focuses on the BAAF model and presents the resulting intensity and polarization patterns.
Finally, Sec.~\ref{sec6} presents a summary and outlook. In this work, we have set the fundamental constants $c$, $G$ to unity, unless otherwise specified, and we will work in the convention $(-,+,+,+)$.

\section{The Konoplya-Zhidenko Spacetime and Photon Sphere}\label{sec2}

We begin by briefly reviewing the Rezzolla–Zhidenko (RZ)
parameterization for a static, spherically symmetric black hole spacetime.
Following the framework originally developed by Rezzolla and Zhidenko \cite{Rezzolla:2014mua} and further refined in \cite{Konoplya:2016jvv},
generic parametric deformations of the Schwarzschild geometry can be
systematically incorporated in a theory-agnostic manner.

The line element of any spherically symmetric and stationary spacetime,
written in spherical coordinates $(t,r,\theta,\phi)$, takes the form
\cite{Rezzolla:2014mua,Konoplya:2016jvv}
\begin{equation}
\df s^2 = -N^2(r)\,\df t^2 + \frac{B^2(r)}{N^2(r)}\,\df r^2
+ r^2 \,\df \theta^2 + r^2 \sin^2\theta \,\df \phi^2 \, .
\end{equation}
Introducing the compactified radial coordinate
\begin{equation}
x \equiv 1 - \frac{r_h}{r} \, ,
\end{equation}
where $r_h$ denotes the event horizon radius, the metric functions can be
expressed as
\begin{equation}
\begin{aligned}
N^2(x) &= x \Big[ 1 - \epsilon (1-x)
+ (a_0 - \epsilon)(1-x)^2
+ \tilde{A}(x)(1-x)^3 \Big] , \\
B(x) &= 1 + b_0 (1-x) + \tilde{B}(x)(1-x)^2 .
\label{metricfunc}
\end{aligned}
\end{equation}
By construction, $x=0$ corresponds to the event horizon, while $x=1$
represents spatial infinity. The functions $\tilde{A}(x)$ and $\tilde{B}(x)$
are introduced through infinite  continued fraction in order to describe the metric near  the horizon and are finite there and at spatial infinity,
\begin{equation}
\tilde{A}(x) = \frac{a_1}{1+\dfrac{a_2 x}{1+\dfrac{a_3 x}{1+\cdots}}} ,
\qquad
\tilde{B}(x) = \frac{b_1}{1+\dfrac{b_2 x}{1+\dfrac{b_3 x}{1+\cdots}}} .
\end{equation}

In this work, we focus on the near-horizon region, where only the lowestorder terms in the expansions are important, such that
$\tilde{A}(x) = a_1$ and $\tilde{B}(x) = b_1$.
Constraints from the parameterized post-Newtonian (PPN) expansion \cite{Will:2014kxa}
fix the parameter
\begin{equation}
\epsilon = -\left(1 - \frac{2M}{r_h}\right),
\end{equation}
where $r_h$ is the event horizon and $M$ is the black hole mass. The coefficients $a_0$ and $b_0$ are tightly bounded to be of order
$10^{-4}$ or smaller \cite{Will:2014kxa,Rezzolla:2014mua}.  Their contributions are therefore
neglected throughout this analysis.

Furthermore, as discussed in \cite{Konoplya:2016jvv}, the parameter $b_1$
controls deformations of the $g_{rr}$ component and do not affects the
effective potential governing particle motion. It is not expected to play
a significant role in the near-horizon matter dynamics relevant to this work. Consequently, we set $b_1 = 0$ in the following.

Under the above assumptions, Eq.~\eqref{metricfunc} can be considerably simplified as
\begin{equation}
N^2(r)
= 1 - \frac{2M}{r}
+ \frac{2M r_h^2 + (a_1 - 1) r_h^3}{r^3}
- \frac{a_1 r_h^4}{r^4} \,,\quad B(r) = 1 \, .
\end{equation}

It is straightforward to verify that the Schwarzschild limit is recovered
when $r_h = 2M$ and $a_1 = 0$.
For $r_h \neq 2M$, the metric function $N^2(r)$ can be interpreted as arising
from parametric deformations of the mass term through a power-series expansion~\cite{Sakalli:2025dif,Magalhaes:2020pyp},
\begin{equation}
M \;\rightarrow\; M + \frac{1}{2}\sum_{i=0}^{\infty}\frac{\eta_i}{r^i} \, .
\end{equation}
To preserve the asymptotic behavior of the Schwarzschild solution at large
distances, the leading coefficient must vanish, $\eta_0 = 0$.
Furthermore, constraints from the PPN framework require $\eta_1 = 0$. Consequently, we restrict our analysis to a single leading-order deformation
of the mass term,
\begin{equation}
M \;\rightarrow\; M + \frac{\eta}{2 r^2} \, ,
\end{equation}
where $\eta$ denotes the Konoplya--Zhidenko (KZ) deformation parameter.
This prescription gives rise to a static spherically symmetric black hole. The corresponding spacetime metric can be written as
\cite{Konoplya:2016pmh,Razzaq:2022esa}
\begin{equation}
\label{metric}
\df s^2
= -\left(1 - \frac{2M}{r} - \frac{\eta}{r^3}\right)\df t^2
+ \frac{1}{1 - \frac{2M}{r} - \frac{\eta}{r^3}}\,\df r^2
+ r^2 \df \theta^2
+ r^2 \sin^2\theta \df \phi^2 \, .
\end{equation}
This line element reduces to the Schwarzschild solution in the limit
$\eta \to 0$, while the $\eta/r^3$ term encodes leading-order deviations
from the Schwarzschild black hole in the strong-field regime.

It is worth emphasizing that the KZ black hole should be
regarded as a parametric deviation from the Schwarzschild geometry, constructed
through a power-series expansion of the mass function, rather than as a solution
derived from a specific modified gravity theory. The deformation term proportional
to $r^{-3}$ represents the lowest-order nontrivial parametric correction to the
Schwarzschild metric that preserves the standard asymptotic behavior at large
distances, while predominantly affecting the strong-field region near the event
horizon. Corrections of this form may arise in effective descriptions motivated by
quantum gravity effects, higher-curvature corrections, or the presence of exotic
matter fields. In this sense, the parameter $\eta$ serves as a
deformation parameter that captures generic departures from the Schwarzschild
geometry, without committing to any particular microscopic mechanism.

The location of the event horizons is determined by the roots of the equation 
\begin{equation}
    g_{tt}=1-\frac{2M}{r}-\frac{\eta}{r^3}=0\,,
\end{equation} 
which gives
\begin{equation}
r_k=\frac{2 M}{3}+\frac{4 M}{3} \cos \left(\frac{\Theta+2 \pi k}{3}\right)\,, \quad k=0\,,1\,,2\,.
\end{equation}
where
\begin{equation}
\Theta=\arccos\left(\frac{16M^3+27\eta}{16M^3}\right)\,.
\end{equation}
In particular, for $\eta>0$, the angle $\Theta$ becomes purely imaginary, and the expression can be rewritten in terms of a hyperbolic cosine, yielding a single positive real root $r_0$ that corresponds to the event horizon. Conversely, for $\eta<-\frac{32}{27} M^3$, the cubic equation admits only one real root, which is negative, indicating the absence of an event horizon and the presence of a naked singularity, thereby violating the cosmic censorship conjecture. For parameter values within the range $-\frac{32}{27} M^3 \le \eta < 0$, the equation admits three real roots, among which $r_0$ and $r_2$ correspond to physical horizons, with $r_0$ representing the event horizon.

Throughout our subsequent analysis, we denote the event horizon radius by $r_h \equiv r_0$ and restrict our investigation to the parameter domain $\eta \ge -\frac{32}{27} M^3$. At the critical value $\eta = -\frac{32}{27} M^3$, the event horizon is located at
$r_h/M = 4/3$, and its radius increases monotonically with increasing $\eta$ \cite{Razzaq:2022esa,Sakalli:2025dif,Magalhaes:2020pyp}. For simplicity, and without loss of generality, we shall set $M = 1$ in the following discussion.

Since the KZ black hole is spherically symmetric, we consider photons moving in the equatorial plane $\theta = \pi/2$. Along a geodesic, a photon possesses two conserved quantities, namely the energy $E = -p_t$ and the angular momentum $L = p_\phi$. Combining these with the null normalization condition $u^\mu u_\mu = 0$, one can obtain the radial equation of motion:
\begin{equation}
    \left(\frac{\df r}{\df \lambda}\right)^2=-V_{\text{eff}}(r,E,L)\,.
\end{equation}
where
\begin{equation}
    V_{\text{eff}}(r,E,L)=E^2\left(1-\frac{1-\frac{2}{r}-\frac{\eta}{r^3}}{r^2}b^2\right)\,.
\end{equation}
Here, we define an impact parameter $b\equiv L/E$. The circular orbits can then be determined by solving the equations $V_{\text{eff}}=0$ and $\partial_r V_{\text{eff}}=0$ with respect to $r$ and $b$. Assuming
a small parameter $\eta$, the radius of the photon sphere $r_{\text{ph}}$ and the critical value of the impact parameter $b_{\text{c}}$ can be expanded
perturbatively in $\eta$ as
\begin{equation}
    \begin{aligned}
        r_{\text{ph}}&=3+\frac{5}{18}\eta-\frac{25}{486}\eta^2+\mathcal{O}(\eta^3)\,,\\
        b_{\text{c}}&=3\sqrt{3}+\frac{\sqrt{3}}{6}\eta-\frac{2\sqrt{3}}{81}\eta^2+\mathcal{O}(\eta^3)\,.
    \end{aligned}
\end{equation}

\begin{figure}[htbp]
	\centering 
    \subfigure[]{\includegraphics[scale=0.42]{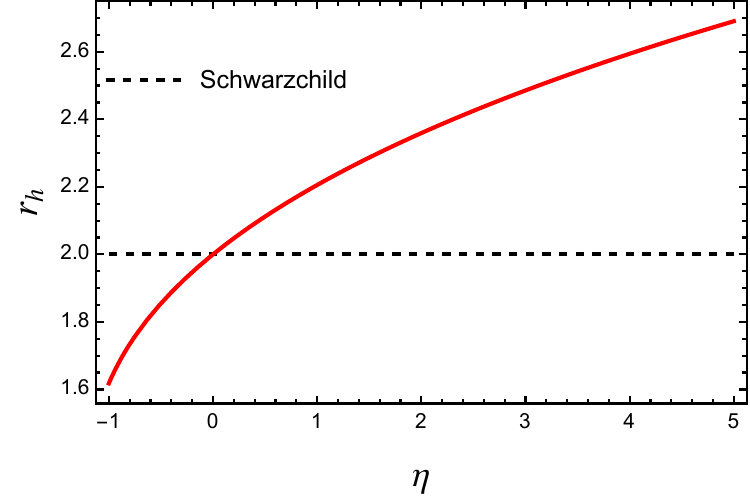}}
	\subfigure[]{\includegraphics[scale=0.42]{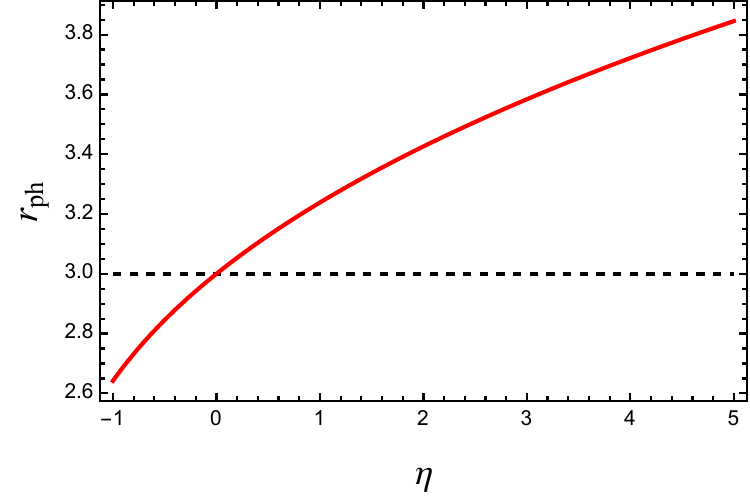}}
	\subfigure[]{\includegraphics[scale=0.42]{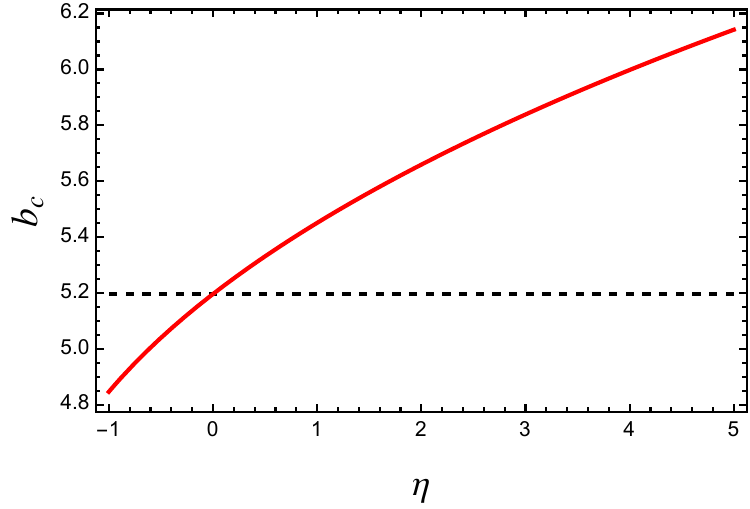}}
    \caption{Dependence of the characteristic quantities of the KZ black hole on the deformation parameter $\eta$: (a) horizon radius $r_h$, (b) photon sphere radius $r_{\text{ph}}$ , and (c) critical impact parameter $b_\text{c}$. The red solid curves correspond to the KZ black hole, while the black dashed lines denote the Schwarzschild case ($\eta=0$).}
 \label{fig:horizon} 
\end{figure}

We compare the analytical perturbative expansion results with the numerical results and find 
that the relative errors of the analytical $r_{\text{ph}}$ for 
$\eta = 1, 2, 3, 4, 5$ are 
$0.37\%$, $2.22\%$, $5.96\%$, $11.66\%$, and $19.31\%$, respectively. 
In addition, for $|\eta|$ values approaching the stability limit 
$\eta = -32/27$, the error is $2.26\%$. Similarly, for $b_{\text{c}}$, the analytical errors for $\eta = 1, 2, 3, 4, 5$ are 
$0.16\%$, $1\%$, $2.76\%$, $5.52\%$, and $9.31\%$, respectively, 
and the error near the stability limit $\eta = -32/27$ is $0.74\%$. We note that the analytical perturbative errors quoted above are only used to assess the validity of the expansion. All black hole images presented in this work are obtained from full numerical ray-tracing calculations based on the exact KZ metric~\eqref{metric}, and therefore are not affected by the perturbative approximation. To illustrate the effect of the deformation parameter $\eta$, we present in Fig.~\ref{fig:horizon} the numerical results for the dependence of several characteristic quantities of the KZ black hole on $\eta$. As shown in panel (a), the horizon radius $r_h$ increases monotonically with $\eta$, deviating from the Schwarzschild value at $\eta = 0$. A similar trend is observed for the photon sphere radius $r_{\text{ph}}$ in panel (b) and for the critical impact parameter $b_\text{c}$ in panel (c). In all cases, larger $\eta$ leads to a more extended spacetime structure, while the Schwarzschild case serves as the reference baseline (black dashed line).
\begin{figure}[htbp]
	\centering 
    \subfigure[]{\includegraphics[scale=0.8]{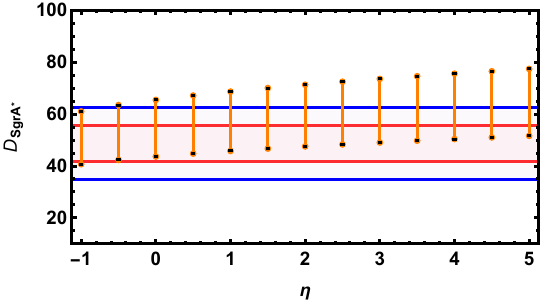}}
	\subfigure[]{\includegraphics[scale=0.8]{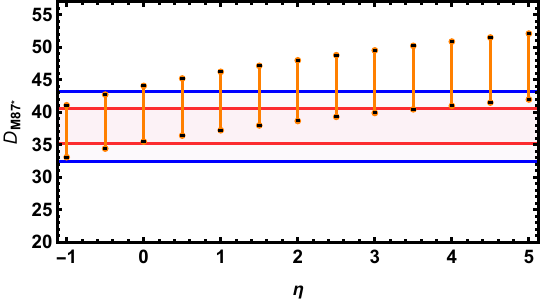}}
    \caption{Comparison of the shadow angular diameter of the KZ black hole with EHT observations for Sgr~A* (left) and M87* (right). The red and blue solid lines denote the $1\sigma$ and $2\sigma$ confidence intervals, respectively. The orange line segments indicate the estimated angular-diameter ranges, with their endpoints marked by thick black ticks.}
 \label{fig:eta} 
\end{figure}
The EHT measurements of the shadow diameters are
$D_{\mathrm{Sgr\,A}^*} = (48.7 \pm 7)\,\mu\mathrm{as}$ for Sgr~A*~\cite{EventHorizonTelescope:2022wkp} and
$D_{\mathrm{M87}^*} = (37.8 \pm 2.7)\,\mu\mathrm{as}$ for M87*~\cite{EventHorizonTelescope:2019dse,EventHorizonTelescope:2019ggy}.
Using these observational constraints, we restrict the deformation parameter $\eta$
and present the corresponding allowed regions in Fig.\ref{fig:eta}.
For both Sgr~A* and M87*, the parameter $\eta$ is constrained to $\eta \lesssim 3$
within the $1\sigma$ confidence interval, and to $\eta \lesssim 5$ within the
$2\sigma$ confidence interval.
To ensure observational relevance, we therefore focus on parameter values within
the $1\sigma$ allowed region throughout this work.

\section{Synchrotron Radiation and Radiative Transfer}
\label{sec3}

We consider the presence of a magnetic field in the vicinity of the black hole, whose explicit configuration will be specified in later sections. In this field, thermal or nonthermal electrons within the accreting plasma undergo synchrotron emission as they are accelerated by the Lorentz force. Our analysis focuses on synchrotron radiation produced by highly relativistic electrons. In this section, we outline the general mechanism of synchrotron emission in the surrounding fluid and describe the subsequent propagation of the emitted radiation from the source to the observer’s image plane. All quantities are expressed in CGS units.

In the plasma, synchrotron radiation is predominantly produced by electrons. 
To correctly describe the emission, absorption, and rotation of polarized radiation in curved spacetime, 
it is essential to construct an appropriate frame of reference. 
Following the formulation of~\cite{Broderick:2003fc}, 
we define the fluid coordinate system by the four-velocity $u^\mu$, the photon wave four-vector $k^\mu$, 
and the local magnetic field $b^\mu$. 
The corresponding orthonormal basis vectors are introduced as follows:
\begin{equation}
e_{(0)}^\mu=u^\mu, \quad e_{(3)}^\mu=\frac{k^\mu}{\omega}-u^\mu, \quad e_{(2)}^\mu=\frac{1}{\mathcal{N}}\left(b^\mu+\beta u^\mu-C e_{(3)}^\mu\right), \quad e_{(1)}^\mu=\frac{\epsilon^{\mu \nu \sigma \rho} u_\nu k_\sigma b_\rho}{\omega \mathcal{N}}\,,
\end{equation}
where $\epsilon^{\mu \nu \sigma \rho}$ is the Levi-Civita tensor, with
\begin{equation}
b^2=b_\mu b^\mu, \quad \beta=u_\mu b^\mu, \quad \omega=-k_\mu u^\mu, \quad C=\frac{k_\mu b^\mu}{\omega}-\beta, \quad \mathcal{N}=\sqrt{b^2+\beta^2-C^2}\,.
\end{equation}
With this choice of frame, all emission, absorption, and Faraday rotation coefficients 
associated with the Stokes parameter $U$ vanish. 
The remaining nonzero emissivities correspond to the Stokes parameters 
$I$, $Q$, and $V$, which are given by 
\cite{Huang:2024bar,Dexter:2016cdk,1980gbs..bookQ....M}:
\begin{equation}
    \begin{aligned}
        j_I&=\frac{\sqrt{3}e^3 B \sin{\theta_B}}{4 \pi m_e c^2}\int^\infty_0 \text{d}\gamma N(\gamma)F(\frac{\nu}{\nu_s})\,,\\
        j_Q&=\frac{\sqrt{3}e^3 B \sin{\theta_B}}{4 \pi m_e c^2}\int^\infty_0 \text{d}\gamma N(\gamma)G(\frac{\nu}{\nu_s})\,,\\
        j_V&=\frac{\sqrt{3}e^3 B \sin{\theta_B}}{4 \pi m_e c^2}\int^\infty_0 \text{d}\gamma N(\gamma)\frac{4\cot{\theta_B}}{3\gamma}H(\frac{\nu}{\nu_s})\,,
    \end{aligned}
\end{equation}
where $N(\gamma)$ denotes the electron energy distribution function, whose form determines the detailed synchrotron emissivity. Here and in what follows, $B$ is the magnitude of the local magnetic field, $\nu$ is the emitted frequency, and $\nu_s = \frac{3 e B \sin{\theta_B}\, \gamma^2}{4\pi m_e c}$ is the characteristic synchrotron frequency. The electron Lorentz factor is $\gamma = 1/\sqrt{1-\beta^2}$, where $e$, $m_e$, and $c$ are the elementary charge, electron mass, and speed of light, respectively. The pitch angle $\theta_B$ is the angle between the wave vector and the magnetic field in the fluid rest frame.

The synchrotron functions corresponding to total, linearly, and circularly polarized emission are defined as
\begin{equation}
  \begin{aligned}
      F(x)&=x\int^\infty_x \text{d}y K_{5/3}(y)\,,\\
      G(x)&=xK_{2/3}(x)\,,\\
      H(x)&=\int^\infty_x \df y K_{1/3}(y)+xK_{1/3}(x)\,,
  \end{aligned}  
\end{equation}
where $K_n(z)$ denotes the modified Bessel function of the second kind of order $n$.

We adopt a relativistic thermal (Maxwellian) electron distribution, which is commonly used for astrophysical synchrotron sources:
\begin{equation}
    N(\gamma)=\frac{n_e \gamma^2 \beta}{\theta_e K_2(1/\theta_e)\text{exp}(-\frac{\gamma}{\theta_e})}\,,
\end{equation}
where $n_e$ is the electron number density and 
$\theta_e = \frac{k_B T_e}{m_e c^2}$ is the dimensionless electron temperature. 
Here $k_B$ is the Boltzmann constant and $T_e$ the thermodynamic temperature. 
In the ultra-relativistic regime ($\beta \approx 1$, $\theta_e \gg 1$), the modified Bessel function can be approximated by $K_2(1/\theta_e) \simeq 2\theta_e^2$. 
Introducing $z \equiv \gamma / \theta_e$, the emissivities become
\begin{equation}
    \begin{aligned}
        j_I&=\frac{\sqrt{3} n_e e^3 B \sin{\theta_B}}{8 \pi m_e c^2}\int^\infty_0 \text{d}z z^2\text{exp}(-z)F(\frac{\nu}{\nu_s})\,,\\
        j_Q&=\frac{\sqrt{3} n_e e^3 B \sin{\theta_B}}{8 \pi m_e c^2}\int^\infty_0 \text{d}z z^2\text{exp}(-z)G(\frac{\nu}{\nu_s})\,,\\
        j_V&=\frac{\sqrt{3} n_e e^3 B \sin{\theta_B}\cot{\theta_B}}{6 \pi m_e \theta_e c^2}\int^\infty_0 \text{d}z z\text{exp}(-z)H(\frac{\nu}{\nu_s})\,.
    \end{aligned}
\end{equation}
Defining $x = (\nu/\nu_s) z^2$, the emissivities can be written compactly as 
\begin{equation}
    \begin{aligned}
        j_I&=\frac{n_e e^2\nu}{2\sqrt{3}c\theta_e^2}I_I(x)\,,\\
        j_Q&=\frac{n_e e^2\nu}{2\sqrt{3}c\theta_e^2}I_Q(x)\,,\\
        j_V&=\frac{2n_e e^2\nu\cot{\theta_B}}{3\sqrt{3}c\theta_e^3}I_V(x)\,,
    \end{aligned}
\end{equation}
where $x \equiv \nu / \nu_c$ is the ratio of the emitted photon frequency to the characteristic frequency of the system $\nu_c = \frac{3 e B \sin{\theta_B} \theta_e^2}{4\pi m_e c}$. 
The dimensionless thermal synchrotron integrals are
\begin{equation}
\begin{aligned}
     I_I(x)&=\frac{1}{x}\int^\infty_0 \df z z^2 \text{exp}(-z) F\left(\frac{x}{z^2}\right)\,,\\
     I_Q(x)&=\frac{1}{x}\int^\infty_0 \df z z^2 \text{exp}(-z) G\left(\frac{x}{z^2}\right)\,,\\
     I_V(x)&=\frac{1}{x}\int^\infty_0 \df z z \text{exp}(-z) H\left(\frac{x}{z^2}\right)\,.
\end{aligned}
\end{equation}

For a hot electron plasma, the absorption coefficients satisfy Kirchhoff’s law,
\begin{equation}
    a_\nu=\frac{j_\nu}{B_\nu}\,,
\end{equation}
where $B_\nu$ represents the Planck black body radiation function. 

Finally, the Faraday rotation coefficients are given by the following expressions:
\begin{equation}
\begin{aligned}
& r_Q=\frac{n_e e^4 B^2 \sin ^2 \theta_B}{4 \pi^2 m_e^3 c^3 \nu^3} f_m(X)+\left(\frac{K_1\left(\theta_e^{-1}\right)}{K_2\left(\theta_e^{-1}\right)}+6 \theta_e\right), \\
& r_V=\frac{n_e e^3 B \cos \theta_B}{\pi m_e^2 c^2 \nu^2} \frac{K_0\left(\theta_e^{-1}\right)-\Delta J_5(X)}{K_2\left(\theta_e^{-1}\right)}\,,
\end{aligned}
\end{equation}
with
\begin{equation}
X=\left(\frac{3}{2 \sqrt{2}} \times 10^{-3} \frac{\nu}{\nu_c}\right)^{-1 / 2}\,.
\end{equation}
Here, $f_m$ and $\Delta J_5$ are fitting functions, and $K_n(X)$ denotes the modified Bessel function of the second kind of order $n$.

Hence, the emissivity, absorption, and Faraday coefficients primarily depend on the electron number density, magnetic-field strength, pitch angle $\theta_B$, and electron temperature.

In order to obtain the image at the observer’s location, we propagate the radiation from the light source to the observer’s screen. To describe the interaction between light rays and matter in radiative transfer, we employ the tensor form of the covariant radiative transfer equation, following \cite{2012ApJ...752..123G}:
\begin{equation}
\label{rteq}
k^\mu \nabla_\mu \mathcal{S}^{\alpha \beta}=\mathcal{J}^{\alpha \beta}+H^{\alpha \beta \mu \nu} \mathcal{S}_{\mu \nu}\,,
\end{equation}
where $\mathcal{S}^{\alpha \beta}$ is the polarization tensor, $k^\mu$ is the photon wave vector, $\mathcal{J}^{\alpha \beta}$ characterizes emission from the source, and $H^{\alpha \beta \mu \nu}$ encodes absorption and Faraday rotation effects. Numerical solutions of the above equation can be obtained using the public code \textsc{Coport}~1.0~\cite{Huang:2024bar}.

Based on \textsc{Coport}~1.0, Huang et al.~\cite{Zhou:2025moa} exploit the gauge invariance of $\mathcal{S}^{\alpha\beta}$ to simplify computations in a suitably chosen parallel-transported frame. In this framework, Eq.~\eqref{rteq} can be decomposed into two parts. The first part encodes gravitational effects:
\begin{equation}
    k^\mu \nabla_\mu f^a=0\,,\,\,f^ak_a=0\,,
\end{equation}
where $f^\mu$ is a normalized spacelike vector orthogonal to $k^\mu$. The second part accounts for the plasma effects:
\begin{equation}
\frac{\df}{\df \lambda} S=R(\chi) J-R(\chi) M R(-\chi) S\,,
\end{equation}
where
\begin{equation}
S=\left(\begin{array}{l}
\mathcal{I} \\
Q \\
\mathcal{U} \\
\mathcal{V}
\end{array}\right), \quad J=\frac{1}{\nu^2}\left(\begin{array}{l}
j_I \\
j_Q \\
j_U \\
j_V
\end{array}\right), \quad M=\nu\left(\begin{array}{cccc}
a_I & a_Q & a_U & a_V \\
a_Q & a_I & r_V & -r_U \\
a_U & -r_V & a_I & r_Q \\
a_V & r_U & -r_Q & a_I
\end{array}\right) \,.
\end{equation}
Here $\mathcal{I}=I/\nu^3$, and similarly for $\mathcal{Q}$, $\mathcal{U}$, and $\mathcal{V}$. The matrix $R(\chi)$ represents the rotation between the synchrotron emission frame and the parallel-transported reference frame:
\begin{equation}
\label{rotationmatrix}
R(\chi)=\left(\begin{array}{cccc}
1 & & & \\
& \cos (2 \chi) & -\sin (2 \chi) & \\
& \sin (2 \chi) & \cos (2 \chi) & \\
& & & 1
\end{array}\right)\,,
\end{equation}
where the rotation angle $\chi$ is defined as the angle between the reference vector $f^\mu$ and the magnetic field $B^\mu$ in the transverse subspace orthogonal to the photon wave vector:
\begin{equation}
\chi=\operatorname{sign}\left(\epsilon_{\mu \nu \alpha \beta} u^\mu f^\nu B^\rho k^\sigma\right) \arccos \left(\frac{P^{\mu \nu} f_\mu B_\nu}{\sqrt{\left(P^{\mu \nu} f_\mu f_\nu\right)\left(P^{\alpha \beta} B_\alpha B_\beta\right)}}\right)\,,
\end{equation}
and $P^{\mu\nu}=g^{\mu\nu}-\frac{k^\mu k^\nu}{\omega^2}+\frac{u^\mu k^\nu+k^\mu u^\nu}{\omega}$ is the induced metric on the transverse subspace.

At the observer’s location, the Stokes parameters are projected onto the image plane using the same rotation matrix~\eqref{rotationmatrix}, with the corresponding rotation angle
\begin{equation}
\chi_o=\operatorname{sign}\left(\epsilon_{\mu \nu \rho \sigma} u_o^\mu f^\nu d^\rho k^\sigma\right) \arccos \left(\frac{P^{\mu \nu} f_\mu d_\nu}{\sqrt{\left(P^{\mu \nu} f_\mu f_\nu\right)\left(P^{\alpha \beta} d_\alpha d_\beta\right)}}\right)\,,
\end{equation}
where $u_o^\mu$ is the observer's four-velocity and $d^\mu$ denotes the $y$-axis direction of the observer’s screen. In this work, we adopt $d^\mu=-\partial_\theta^\mu$. The projected Stokes parameters are then
\begin{equation}
\mathcal{I}_o=\mathcal{I}, \quad \mathcal{Q}_o=\mathcal{Q} \cos 2\chi_o-\mathcal{U} \sin 2\chi_o, \quad \mathcal{U}_o=\mathcal{Q} \sin 2\chi_o+\mathcal{U} \cos 2\chi_o, \quad \mathcal{V}_o=\mathcal{V}\,.
\end{equation}

The observed Stokes parameters encode the polarization state of the radiation. The total intensity is given by $\mathcal{I}_o$, while $\mathcal{V}_o$ describes circular polarization — positive (negative) values correspond to left- (right-) handed circular polarization. The linear polarization intensity and electric-vector position angle (EVPA) are given by
\begin{equation}
P_o=\sqrt{\mathcal{Q}_o^2+\mathcal{U}_o^2}, \quad \Phi_{\mathrm{EVPA}}=\frac{1}{2} \arctan \frac{\mathcal{U}_o}{\mathcal{Q}_o}\,.
\end{equation}

In the following, we investigate the image features of the KZ black hole using two representative models of geometrically thick accretion flows. The first is a phenomenological RIAF model, while the second is an analytic thick-disk model proposed by Hou \textit{et al.} \cite{Hou:2023bep}. For convenience, we shall refer to the latter as the ballistic approximation accretion flow model (hereafter BAAF).

\section{RIAF Model}
\label{sec4}

We work in cylindrical coordinates, where 
$R = r \sin\theta$ denotes the cylindrical radius and $z = r \cos\theta$ measures the vertical height from the equatorial plane at $\theta = \pi/2$. 
Following the analytic construction of RIAF models~\cite{Broderick:2010kx}, we prescribe the radial and vertical profiles of the number density and temperature distributions as
\begin{equation}
n_e=n_h\left(\frac{r}{r_h}\right)^{-2} \exp \left(-\frac{z^2}{2(\alpha R)^2}\right), \quad T_e=T_h\left(\frac{r}{r_h}\right)^{-1}\,,
\end{equation}
where $r_h$ is the outer horizon radius, and $n_h$ and $T_h$ denote the electron number density and temperature at the horizon, respectively. 
The parameter $\alpha$ is a dimensionless constant controlling the disk thickness. Since we do not explore the impact of disk thickness on the imaging in this work, we simply adopt a representative choice for a thick disk~\cite{Vincent:2022fwj}, $\alpha=1$. It is worth noting that reasonable variations of the power-law indices in the density and temperature profiles mainly affect the overall brightness distribution and image contrast, while leaving the photon ring location and shadow size essentially unchanged. Consequently, the imaging signatures associated with the deformation parameter $\eta$ are robust against such variations of the RIAF power-law indices. Accordingly, we adopt the representative exponents $-2$ and $-1$ following standard RIAF phenomenology.

The local magnetic field strength is characterized by the cold magnetization parameter $\sigma$, defined as
\begin{equation}
B=\sqrt{\sigma\rho}\,,
\end{equation}
where $\rho = n_e m_p c^2$ is the fluid mass density. 
In general, $\sigma$ can be modeled as a spatially varying distribution, e.g., $\sigma = \sigma_h \, r_h/r$~\cite{Pu:2016qak}, where $\sigma_h$ is its value at the event horizon. 
For simplicity, we assume $\sigma$ to be constant and fix it to $\sigma = 0.1$ in all models considered below.

The fluid dynamics in this model is relatively unconstrained. We therefore consider three representative kinematic configurations:

\paragraph*{(a) Orbiting motion} 
For circular motion around the black hole~\cite{Gold:2020iql}, the fluid four-velocity has only $u^t$ and $u^\phi$ components, given by
\begin{equation}
u^\mu = u^t (1, 0, 0, \Omega)\,,
\end{equation}
where 
\begin{equation}
u^t = \sqrt{-\frac{1}{g_{tt} + g_{\phi\phi}\Omega^2}}, 
\qquad
\Omega = \frac{u^\phi}{u^t} = \frac{g_{tt} l}{g_{\phi\phi}}, 
\qquad
l = -\frac{u_\phi}{u_t} = l_0 \frac{R^{3/2}}{C + R}, 
\qquad
R = r \sin\theta\,,
\end{equation}
Here, $l$ denotes the specific angular momentum, and $l_0$ and $C$ are free parameters.
Following Ref.~\cite{Gold:2020iql}, we set $l_0 = C = 1$ throughout this work.

\paragraph*{(b) Infalling motion}
Assuming that the fluid is at rest at spatial infinity, i.e., $u_t = -1$, the four-velocity is given by
\begin{equation}
\label{infalling}
u^\mu = (-g^{tt},\, \sqrt{-(1 + g^{tt}) g^{rr}},\, 0,\, 0)\,,
\end{equation}
which corresponds to a special case of the conical solution.

\paragraph*{(c) Combined motion}
A more general flow configuration can be constructed as a combination of orbiting and infalling motion~\cite{Pu:2016qak}, with
\begin{equation}
u^\mu = (u^t,\, u^r,\, 0,\, u^\phi)\,,
\end{equation}
where
\begin{equation}
\Omega = \frac{u^\phi}{u^t} = \Omega_o + \beta_1(\Omega_i - \Omega_o), 
\qquad
u^r = u_o^r + \beta_2 (u_i^r - u_o^r)\,,
\end{equation}
and
\begin{equation}
\Omega_o = -\frac{g_{tt} l}{g_{\phi\phi}}, 
\qquad 
\Omega_i = 0, 
\qquad 
u_o^r = 0, 
\qquad 
u_i^r = \sqrt{-(1 + g^{tt}) g^{rr}}\,.
\end{equation}
Here, the subscripts “$i$” and “$o$” refer to the infalling and orbiting components, respectively. 
The parameters $\beta_1$ and $\beta_2$ are adjustable coefficients in the range $(0,1)$, controlling the relative weights of the two components: smaller values correspond to predominantly orbiting motion, while larger values indicate a more infalling flow. 
Throughout this work, we adopt $\beta_1 = \beta_2 = 0.2$. The temporal component 
$u^t$ is then determined by $u^2=-1$.
\begin{equation}
u^t=\sqrt{-\frac{1+g_{r r}\left(u^r\right)^2}{g_{t t}+\Omega^2 g_{\phi \phi}}}\,.
\end{equation}

Next, let us consider the magnetic field. Under the assumption of ideal MHD,
\begin{equation}
    u_{\mu}F^{\mu\nu}=0\,.
\end{equation}
the electric field in the rest frame of the fluid, $e^\nu = u_\mu F^{\mu\nu} = 0$, and the magnetic field 
$b^\nu = u_\mu {^*F}^{\mu\nu}$ is orthogonal to the fluid four-velocity, $b_\mu u^\mu = 0$. In realistic accretion flows, the predominant plasma motion is differential rotation, which naturally winds the magnetic field into a predominantly toroidal configuration~\cite{Komissarov:2006nz,Oda:2009am}. This behavior is also strongly supported by a wide range of numerical studies~\cite{Saurabh:2025kwb,Broderick:2010kx,Hawley:1999xv,Singh:2024asb}, where the disk region is overwhelmingly toroidal field dominated. In addition, a purely toroidal magnetic field is sufficient to generate the large-scale poloidal magnetic flux required for launching jets~\cite{Liska:2018btr}. In this work, we focus on the pitch angle on the appearance of thick-disk images, thus for simplicity and clarity, we assume a purely toroidal magnetic field~\cite{Komissarov:2006nz},
\begin{equation}
\label{toroidalmag}
    b^\mu\sim(l,0,0,1)\,,
\end{equation}
We emphasize that, although we consider a purely toroidal magnetic field in our analysis, real accretion flows exhibit complex magnetic field topologies, with both toroidal and poloidal components. These mixed configurations can modify the pitch angle, which in turn affects the intensity distribution in the resulting black hole images. In particular, poloidal magnetic components can break the symmetry in the emission patterns, leading to variations in the observed brightness, as discussed in the following figures in Sec.\ref{ani}. Therefore, the purely toroidal model serves as a simplified approximation, and more realistic field configurations may introduce additional asymmetries in the brightness distribution.

Depending on whether the emissivity depends on the angle between the magnetic field and the emitted photons, we distinguish two cases: isotropic and anisotropic emission. In the isotropic case, the emissivity is independent of this angle, and only the magnetic field strength needs to be specified. We employ  an angle-averaged emissivity defined as
    
\begin{equation}
\bar{j}_\nu=\frac{1}{2} \int_0^\pi j_\nu \sin \theta_B \mathrm{d}\theta_B\,,
\end{equation}
with the fitting formula from ~\cite{Mahadevan:1996cc}:
\begin{equation}
\bar{j}_\nu=\frac{n_e e^2 \nu}{2 \sqrt{3} c \theta_e^2} I(x), \quad x=\frac{\nu}{\nu_c}, \quad \nu_c=\frac{3 e B \theta_e^2}{4 \pi m_e c}\,,
\end{equation}
where the dimensionless function $I(x)$ is approximated as~\cite{Mahadevan:1996cc}
\begin{equation}
I(x)=\frac{4.0505}{x^{1 / 6}}\left(1+\frac{0.4}{x^{1 / 4}}+\frac{0.5316}{x^{1 / 2}}\right) \exp \left(-1.8899 x^{1 / 3}\right)\,.
\end{equation}
For the anisotropic case, the direction of the magnetic field must be accounted for. 
Adopting the toroidal field configuration in Eq.~\eqref{toroidalmag}, the corresponding fitting function is
\begin{equation}
I(x)=2.5651\left(1+1.92 x^{-1 / 3}+0.9977 x^{-2 / 3}\right) \exp \left(-1.8899 x^{1 / 3}\right)\,.
\end{equation}

In the next section, we present the black hole images obtained from the phenomenological RIAF model, based on the methodology described above.

\subsection{Isotropic Radiation Case}

\begin{figure}[htbp]
	\centering 
    \subfigure[$\eta=-1,\theta_o=0.001^\circ$]{\includegraphics[scale=0.45]{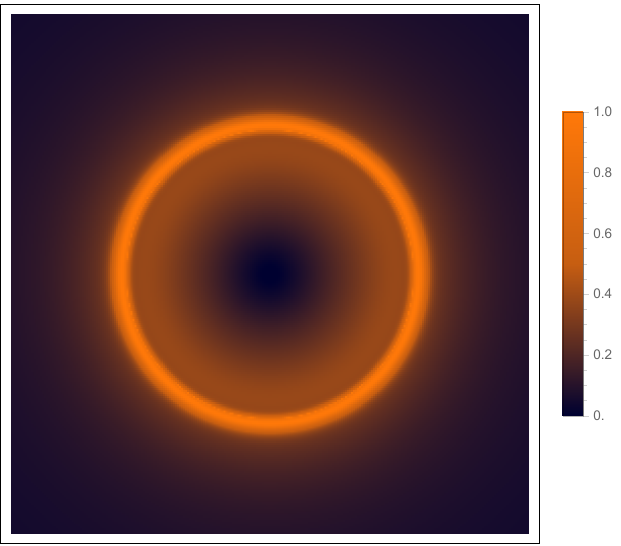}}
	\subfigure[$\eta=0.1,\theta_o=0.001^\circ$]{\includegraphics[scale=0.45]{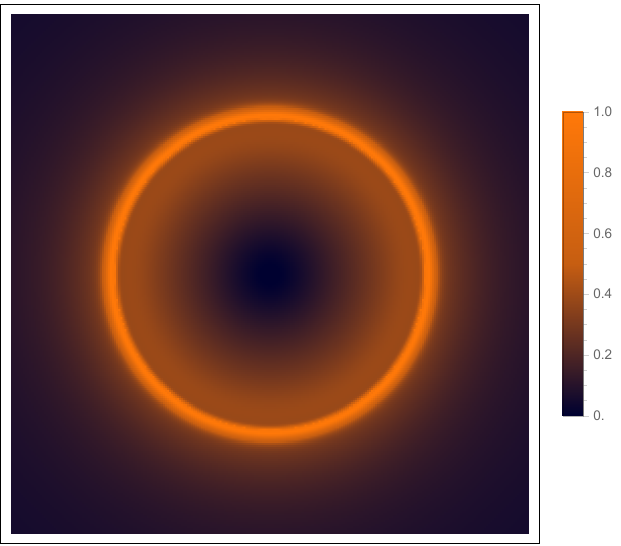}}
	\subfigure[$\eta=2,\theta_o=0.001^\circ$]{\includegraphics[scale=0.45]{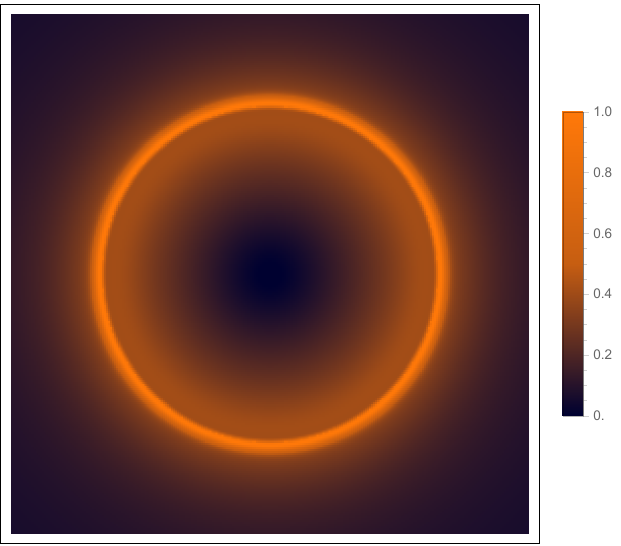}}

\subfigure[$\eta=-1,\theta_o=17^\circ$]{\includegraphics[scale=0.45]{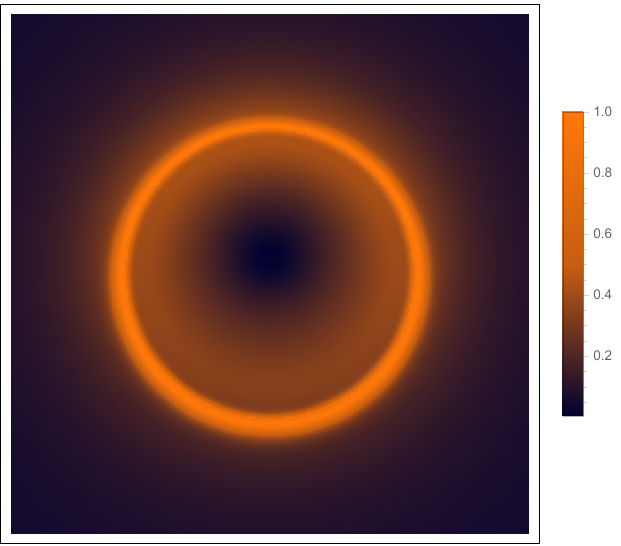}}
	\subfigure[$\eta=0.1,\theta_o=17^\circ$]{\includegraphics[scale=0.45]{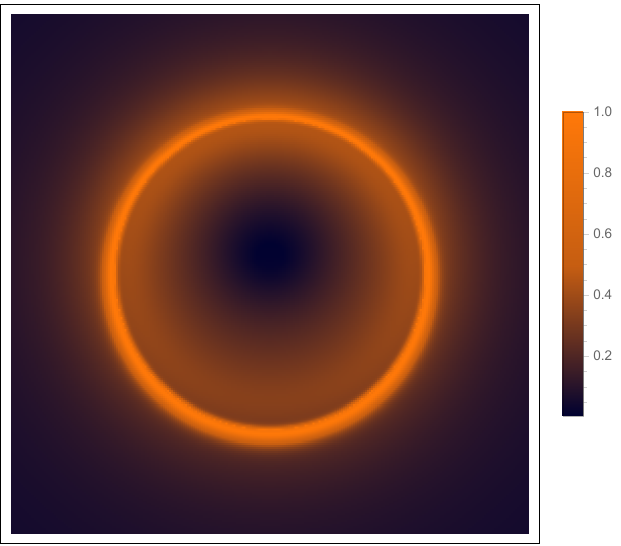}}
	\subfigure[$\eta=2,\theta_o=17^\circ$]{\includegraphics[scale=0.45]{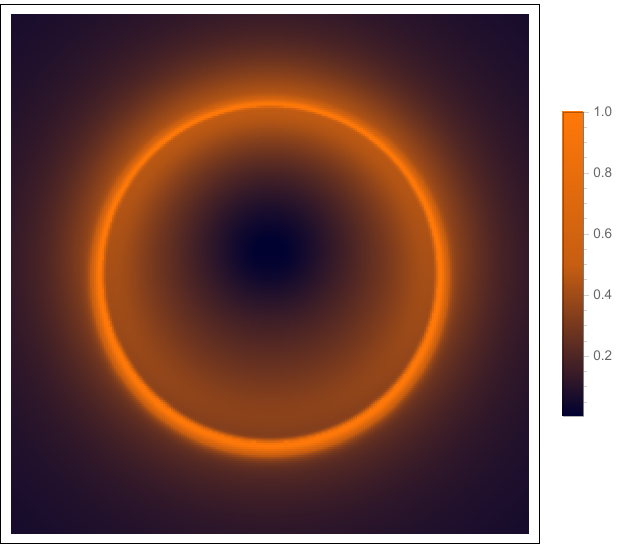}}
    
    \subfigure[$\eta=-1,\theta_o=60^\circ$]{\includegraphics[scale=0.45]{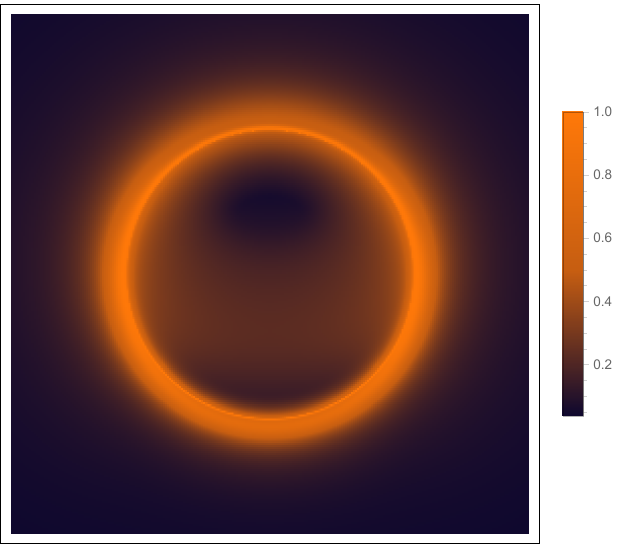}}
	\subfigure[$\eta=0.1,\theta_o=60^\circ$]{\includegraphics[scale=0.45]{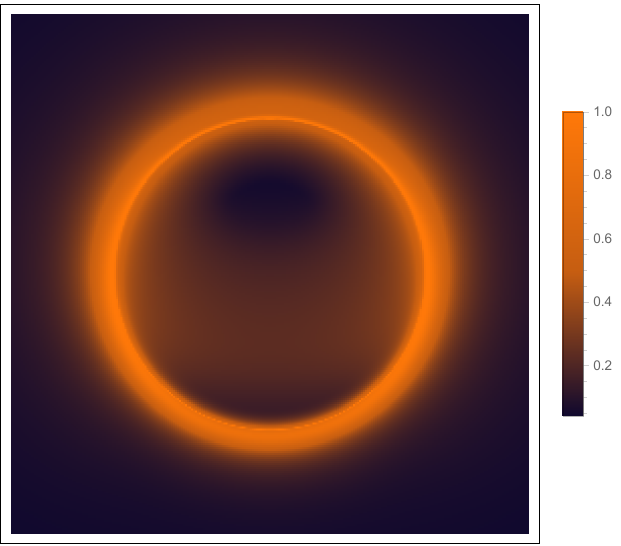}}
	\subfigure[$\eta=2,\theta_o=60^\circ$]{\includegraphics[scale=0.45]{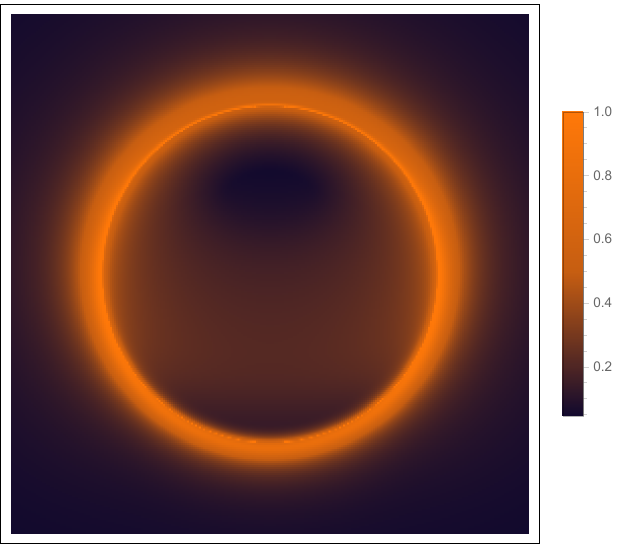}}

     \subfigure[$\eta=-1,\theta_o=85^\circ$]{\includegraphics[scale=0.45]{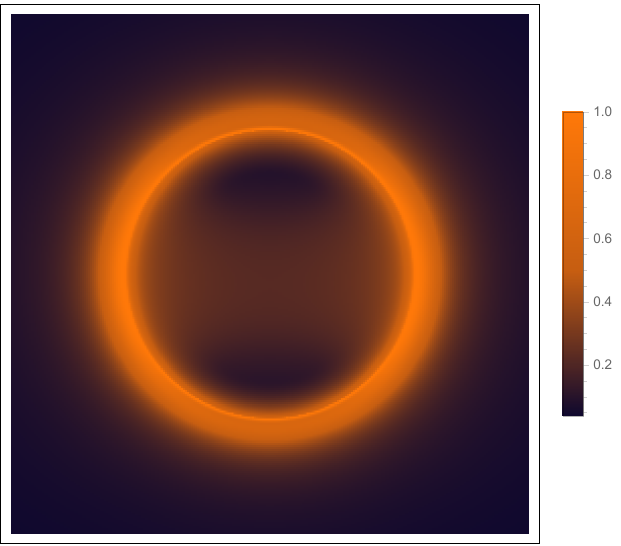}}
	\subfigure[$\eta=0.1,\theta_o=85^\circ$]{\includegraphics[scale=0.45]{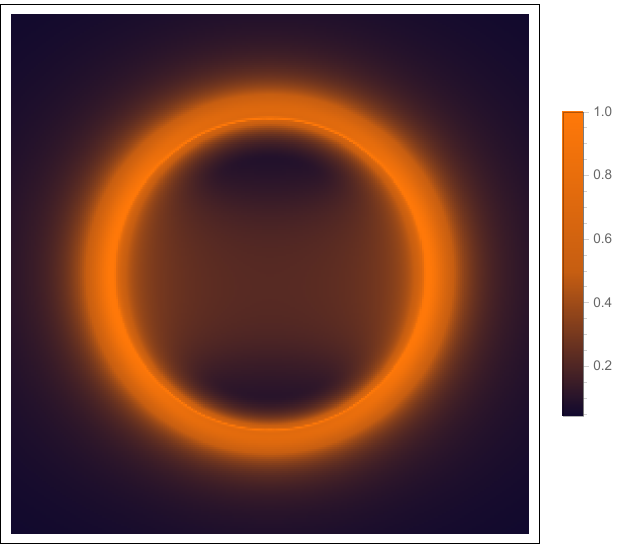}}
	\subfigure[$\eta=2,\theta_o=85^\circ$]{\includegraphics[scale=0.45]{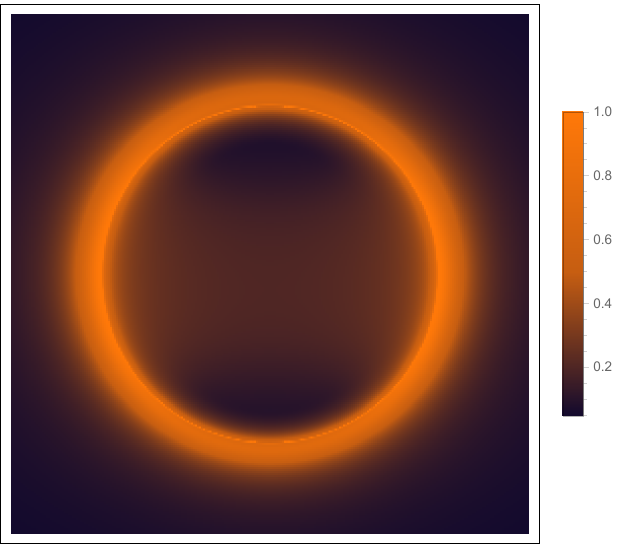}}
    \caption{Intensity maps of the KZ black hole in the RIAF model with isotropic emission. The accretion flow follows the infalling motion, the observing frequency is 230 GHz.}
 \label{fig:1} 
\end{figure}
\begin{figure}[htbp]
	\centering

    \subfigure[$\theta_o=0.001^\circ$]{\includegraphics[scale=0.85]{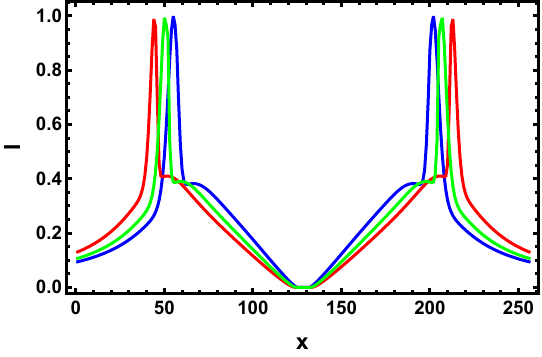}}
	\subfigure[$\theta_o=17^\circ$]{\includegraphics[scale=0.85]{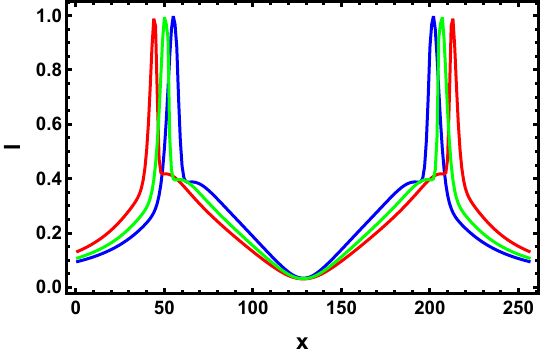}}
	\subfigure[$\theta_o=60^\circ$]{\includegraphics[scale=0.85]{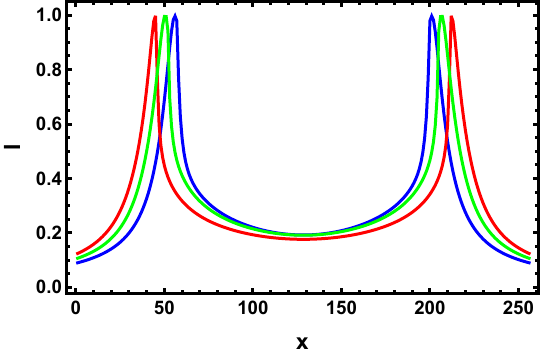}}
    \subfigure[$\theta_o=85^\circ$]{\includegraphics[scale=0.85]{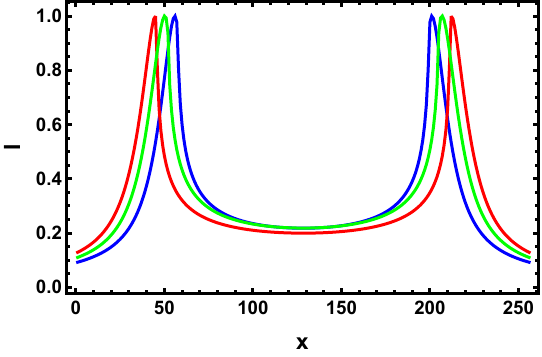}}

    \caption{
     Intensity distribution along the screen $x$-axis for the KZ black hole in the RIAF model with isotropic emission. 
     The accretion flow follows the infalling motion, and the observing frequency is fixed at $230\,\mathrm{GHz}$. 
     The curves correspond to different values of $\eta$: red for $\eta=2$, green for $\eta=0.1$, and blue for $\eta=-1$.
     }
    \label{fig:2}
\end{figure}

\begin{figure}[htbp]
	\centering 
\subfigure[$\theta_o=0.001^\circ$]{\includegraphics[scale=0.85]{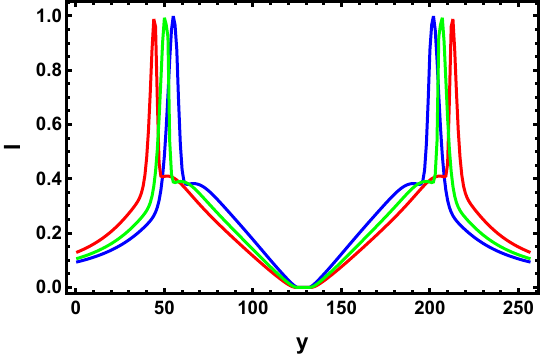}}
	\subfigure[$\theta_o=17^\circ$]{\includegraphics[scale=0.85]{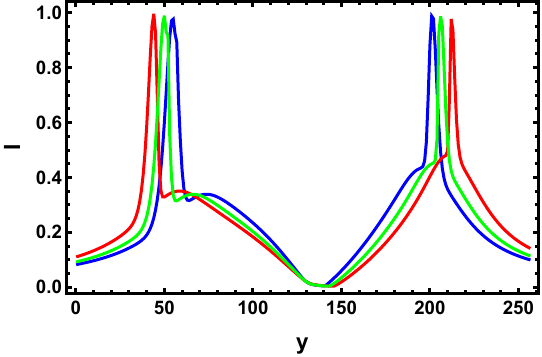}}
	\subfigure[$\theta_o=60^\circ$]{\includegraphics[scale=0.85]{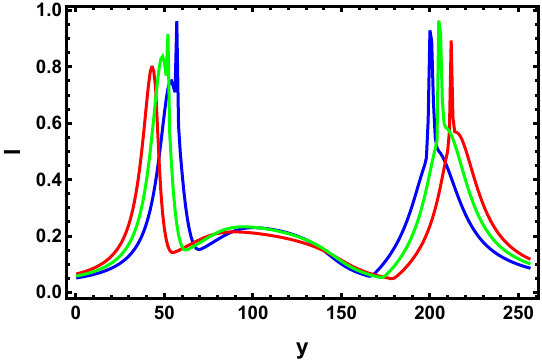}}
    \subfigure[$\theta_o=85^\circ$]{\includegraphics[scale=0.85]{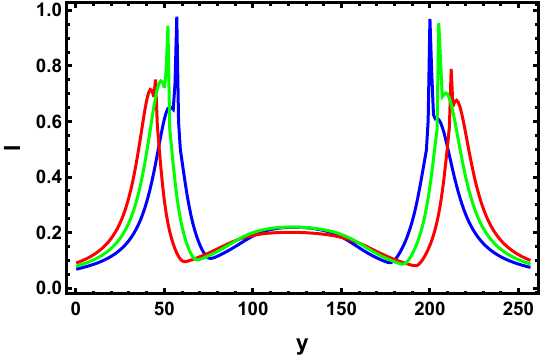}}

    \caption{Intensity distribution along the screen $y$-axis for the KZ black hole in the RIAF model with isotropic emission. 
     The accretion flow follows the infalling motion, and the observing frequency is fixed at $230\,\mathrm{GHz}$. 
     The curves correspond to different values of $\eta$: red for $\eta=2$, green for $\eta=0.1$, and blue for $\eta=-1$.}
    \label{fig:3}
\end{figure}
Fig.~\ref{fig:1} shows the intensity maps of the KZ black hole in the RIAF model with isotropic emission.
The accretion flow follows an infalling motion, and the observing frequency is fixed at $230\mathrm{GHz}$.
Each panel corresponds to a different choice of the deformation parameter $\eta$ and the observer’s inclination angle $\theta_o$:
from top to bottom, the inclination varies as $\theta_o = 0.001^\circ, 17^\circ, 60^\circ, 85^\circ$, while from left to right, the deformation parameter takes the values $\eta=-1, 0.1, 2$.
For a more quantitative comparison, we also analyze the intensity distributions along the screen’s $x$- and $y$-axes in Fig.~\ref{fig:2} and Fig.~\ref{fig:3}, where $\eta=2$, $\eta=0.1$, and $\eta=-1$ are shown in red, green, and blue, respectively.

The panels in Fig.~\ref{fig:1} consistently exhibit a pronounced bright ring, corresponding to the peaks seen in Figs.~\ref{fig:2} and \ref{fig:3}.
This feature originates from higher-order images, namely photons that orbit the black hole one or more times before reaching the observer, and is a direct consequence of strong gravitational lensing.
Outside the ring lies the primary image, formed by photons reaching the observer directly from the disk.
The dark regions in the images correspond to the event horizon.
For geometrically thin disks, the horizon produces a distinct inner shadow that could be detected by the EHT~\cite{Chael:2021rjo}.
In contrast, for geometrically thick disks, emission from off-equatorial regions obscures the horizon boundary, making the inner shadow harder to observe.

By comparing different subplots, we can examine the effects of the deformation parameter $\eta$ and the inclination angle $\theta_o$ on the black hole image.
For fixed $\theta_o$, increasing $\eta$ enlarges both the bright ring and the central dark region without altering their shapes, since the horizon radius $r_h$ and the critical impact parameter $b_\text{c}$ of the photon sphere both grow with $\eta$, as discussed in Sec.~\ref{sec2}.
For fixed $\eta$, varying $\theta_o$ produces the following trends:
for polar viewing, the bright ring (higher-order images) and the dark region remain centered and isotropic;
at $\theta_o=17^\circ$, a clear up–down asymmetry in the bright ring emerges;
at $\theta_o=60^\circ$, two distinct dark regions appear inside the bright ring, with the upper one slightly dimmer;
and at $\theta_o=85^\circ$, the image becomes nearly symmetric, though the left–right brightness still exceeds the up–down one.
The persistent left–right brightness symmetry arises from the spherical symmetry of the spacetime and the choice of infalling accretion flow.
In contrast, the up–down brightness dependence reflects the equatorial symmetry of the thick disk: for observers near the equatorial plane, high-latitude emission partially fills the dark region, while near the poles, insufficient photons reach the observer.
Figs.~\ref{fig:2} and \ref{fig:3} further illustrate these effects through horizontal and vertical intensity cuts.

\begin{figure}[htbp]
	\centering

    \subfigure[$85$ $\mathrm{GHz}$]{\includegraphics[scale=0.45]{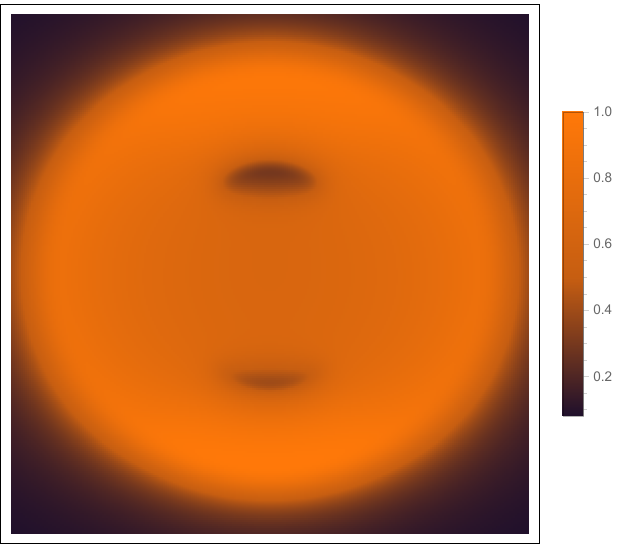}}
	\subfigure[$230$ $\mathrm{GHz}$]{\includegraphics[scale=0.45]{theta=85du,eta=-1_tong.pdf}}
    \subfigure[$345$ $\mathrm{GHz}$]{\includegraphics[scale=0.45]{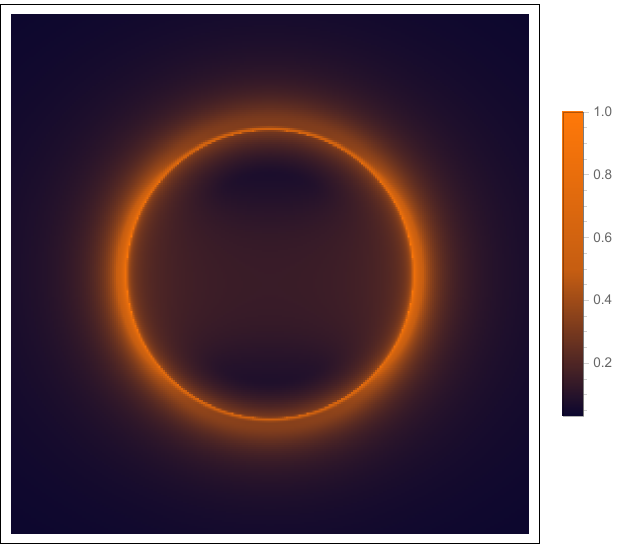}}

    \subfigure[]{\includegraphics[scale=0.6]{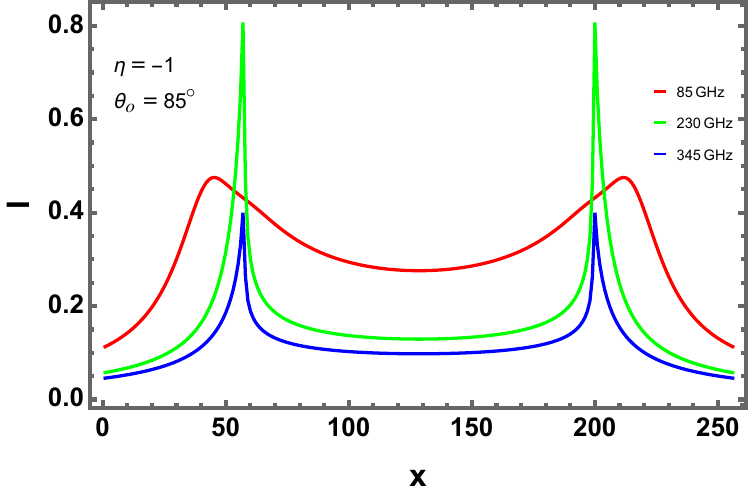}}
    \subfigure[]{\includegraphics[scale=0.6]{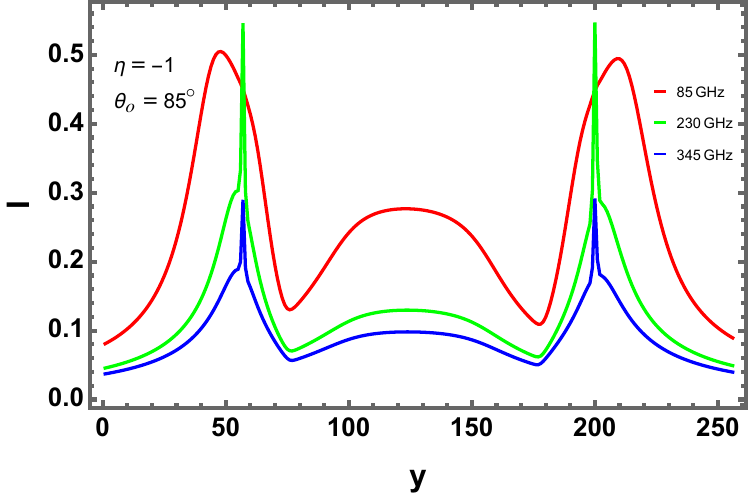}}

    \caption{Intensity maps of the KZ black hole in the RIAF model with isotropic emission at different observing frequencies. The accretion flow follows the infalling motion, with fixed parameters $\theta_o = 85^\circ$ and $\eta = -1$. 
    Panels (a)–(c) show the images at observing frequencies of $85\,\mathrm{GHz}$, $230\,\mathrm{GHz}$, and $345\,\mathrm{GHz}$, respectively. 
    Panels (d) and (e) display the corresponding intensity profiles along the horizontal ($x$-axis) and vertical ($y$-axis) cuts through the image center, where the curves for different frequencies are color-coded: red for $85\,\mathrm{GHz}$, green for $230\,\mathrm{GHz}$, and blue for $345\,\mathrm{GHz}$.
    }
    \label{fig:5}
\end{figure}

In Fig.~\ref{fig:5}, the dependence of the KZ black hole image on observing frequency is illustrated for the RIAF model with isotropic emission.
For fixed parameters $\theta_o=85^\circ$ and $\eta=-1$, the image transitions from a diffuse morphology at $85\,\mathrm{GHz}$ to a well-defined ring at $230\,\mathrm{GHz}$ and $345\,\mathrm{GHz}$.
At the lowest frequency, the dark region is almost entirely hidden, and the distinction between the primary and higher-order features becomes unclear.
With increasing frequency, the contribution from regions far outside the photon ring is strongly suppressed, resulting in a more pronounced visibility of both the higher-order structures and the dark region boundary.
The corresponding horizontal and vertical intensity cuts confirm this trend: higher observing frequencies enhance the contrast of the photon ring while diminishing the diffuse background contribution. 

\begin{figure}[htbp]
	\centering

    \subfigure[Orbiting motion]{\includegraphics[scale=0.45]{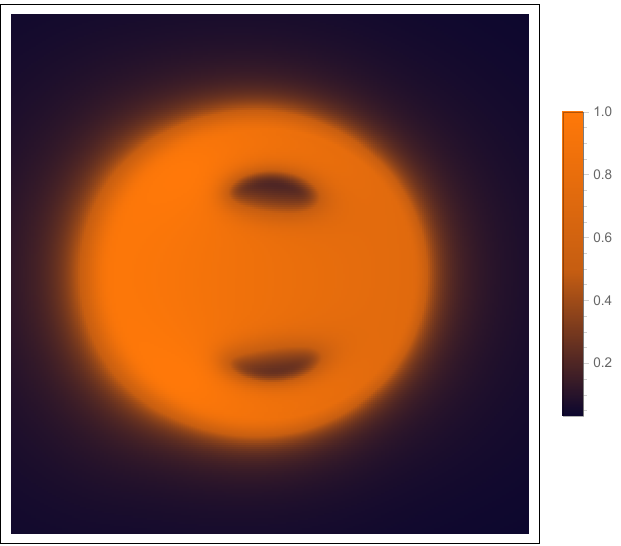}}
	\subfigure[Infalling motion]{\includegraphics[scale=0.45]{theta=85du,eta=-1_tong.pdf}}
    \subfigure[Combined motion]{\includegraphics[scale=0.45]{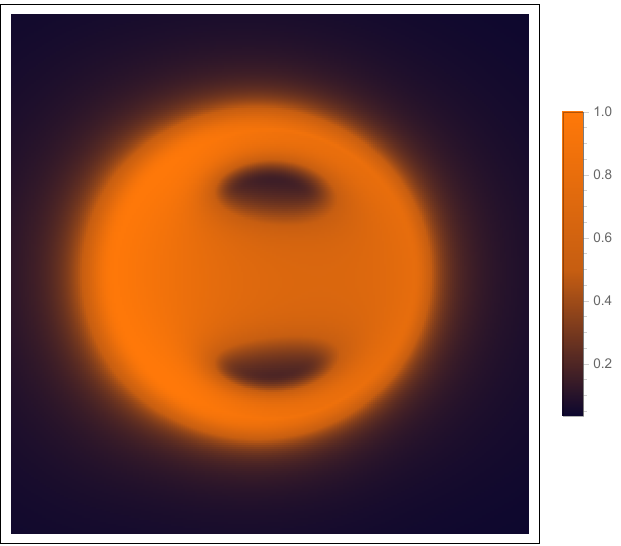}}

    \subfigure[]{\includegraphics[scale=0.5]{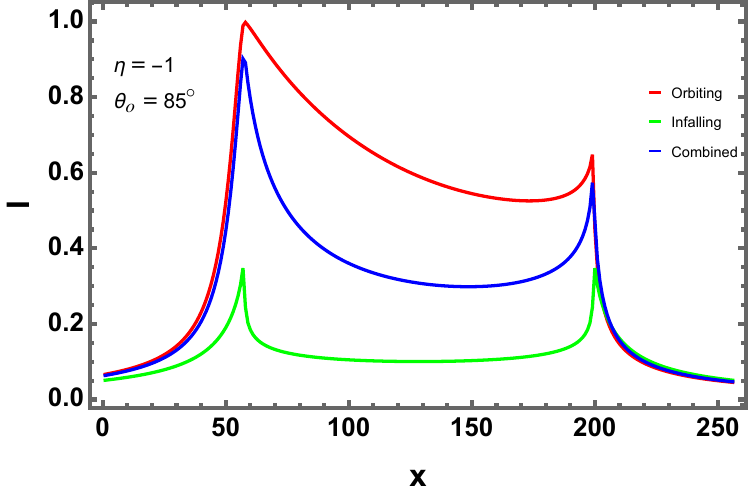}}
    \subfigure[]{\includegraphics[scale=0.5]{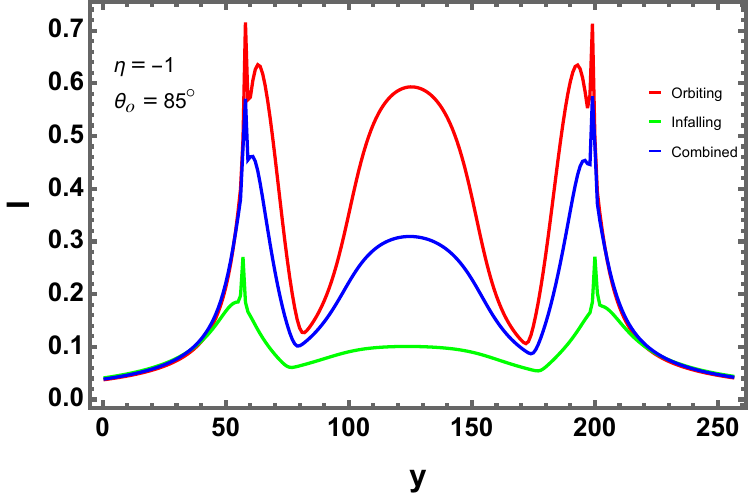}}

    \caption{Intensity maps of the KZ black hole in the RIAF model with isotropic emission at different flow dynamics. Other parameters are set to $\nu_o=230\,\mathrm{GHz}$, $\theta = 85^\circ$ and $\eta = -1$. 
    Panels (a)–(c) show the images with flow of orbiting, infalling, and combined motion, respectively. 
    Panels (d) and (e) display the corresponding intensity profiles along the horizontal ($x$-axis) and vertical ($y$-axis) cuts through the image center, where the curves for different frequencies are color-coded: red for orbiting motion, green for infalling motion, and blue for combined motion.}
    \label{fig:6}
\end{figure}

In Fig.~\ref{fig:6}, we further compare the effects of different accretion-flow dynamics on the image morphology in the same model. The observing frequency is fixed at $\nu_o = 230\,\mathrm{GHz}$, with parameters $\theta_o = 85^\circ$ and $\eta = -1$. Panels (a)–(c) correspond to the orbiting, infalling, and combined motion, respectively, illustrating how the image morphology depends on the flow pattern.
Panels (d) and (e) show the corresponding horizontal ($x$-axis) and vertical ($y$-axis) intensity cuts through the image center, where red, green, and blue denote the orbiting, infalling, and combined cases, respectively.
Since we set $\beta_1 = \beta_2 = 0.2$, the orbiting component dominates the combined flow, explaining the close similarity between the orbiting and combined images.
The orbiting motion tends to obscure the dark region by enhancing off-equatorial emission and substantially increasing the total brightness, whereas the radial inflow produces more distinct direct and higher-order features.
This difference originates from their respective Doppler patterns: for a purely radial inflow, the Doppler shifts remain nearly uniform in all directions, yielding a relatively sharp image; in contrast, a purely azimuthal flow produces strong redshifts and blueshifts in different azimuthal sectors.
Gravitational lensing subsequently maps these sectors onto overlapping regions of the observer’s sky, such that photons with very different frequency shifts contribute to the same pixel, effectively smearing the image.

\subsection{Anisotropic Radiation Case}
\label{ani}

In the previous subsection, we examined the imaging features of the KZ black hole in the RIAF thick-disk model under isotropic synchrotron emission.
It is important to note, however, that synchrotron radiation is intrinsically anisotropic—its emissivity depends sensitively on the emission direction.
This dependence is governed by the pitch angle $\theta_B$, defined as the angle between the photon wave vector and the magnetic field in the fluid rest frame.
In this subsection, we extend our analysis to account for this anisotropy, assuming a toroidal magnetic field as specified in Eq.~\eqref{toroidalmag}.
\begin{figure}[htbp]
	\centering 
	\subfigure[$\eta=-1,\theta_o=0.001^\circ$]{\includegraphics[scale=0.45]{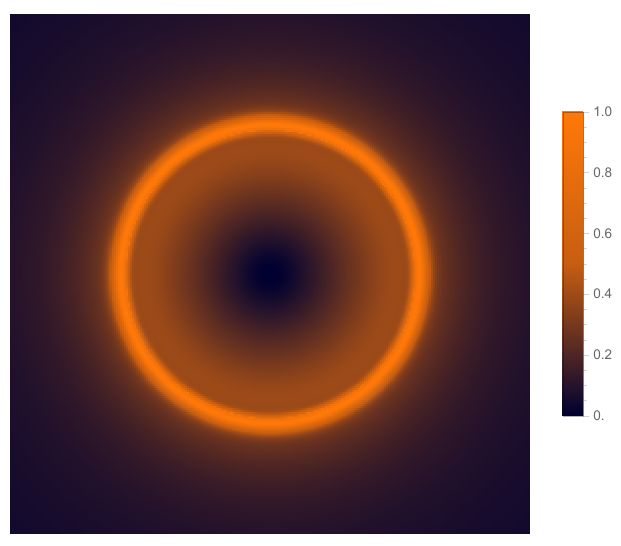}}
	\subfigure[$\eta=0.1,\theta_o=0.001^\circ$]{\includegraphics[scale=0.45]{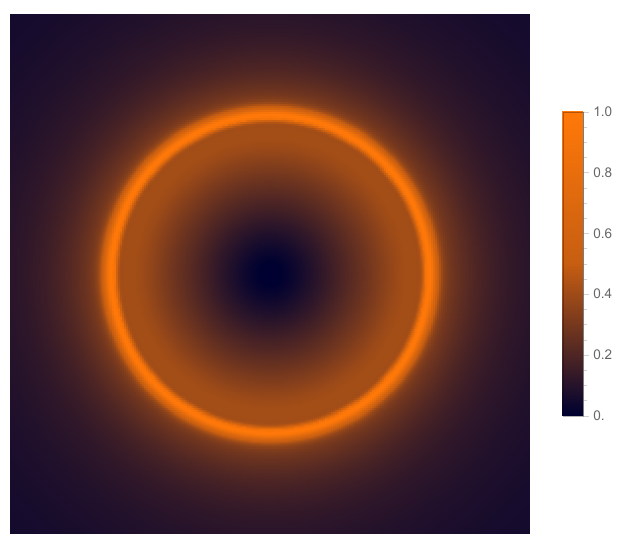}}
	\subfigure[$\eta=2,\theta_o=0.001^\circ$]{\includegraphics[scale=0.45]{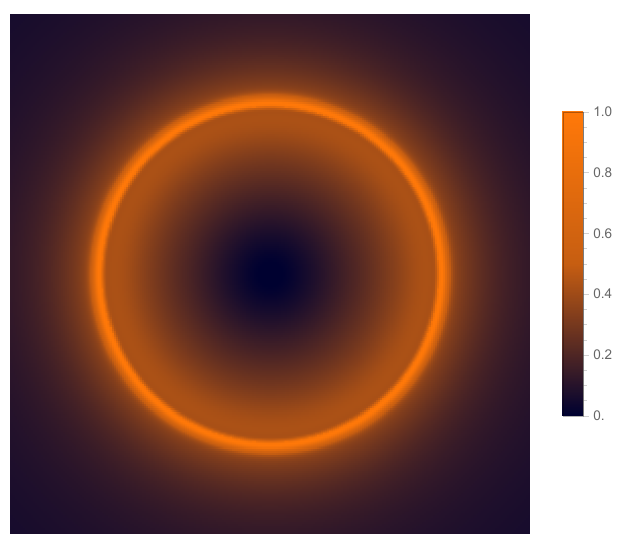}}

\subfigure[$\eta=-1,\theta_o=17^\circ$]{\includegraphics[scale=0.45]{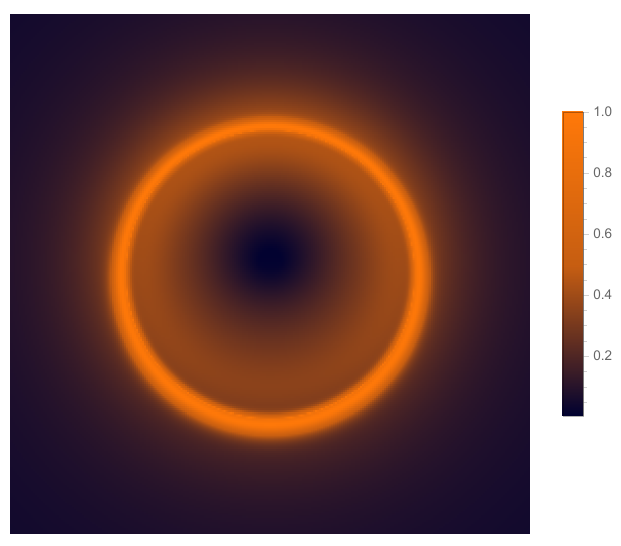}}
	\subfigure[$\eta=0.1,\theta_o=17^\circ$]{\includegraphics[scale=0.45]{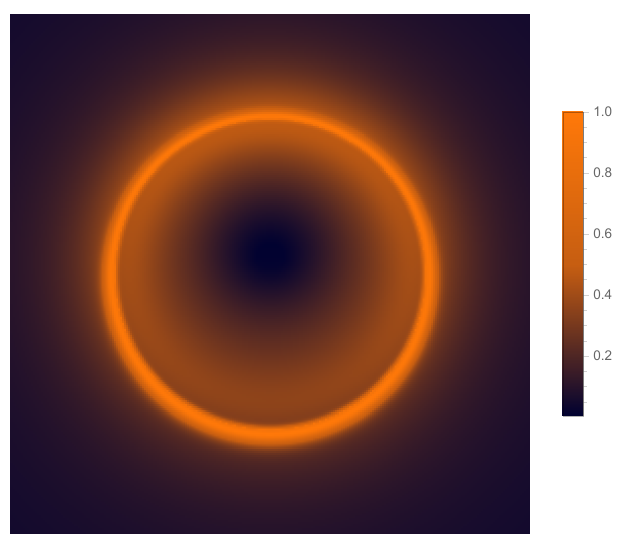}}
	\subfigure[$\eta=2,\theta_o=17^\circ$]{\includegraphics[scale=0.45]{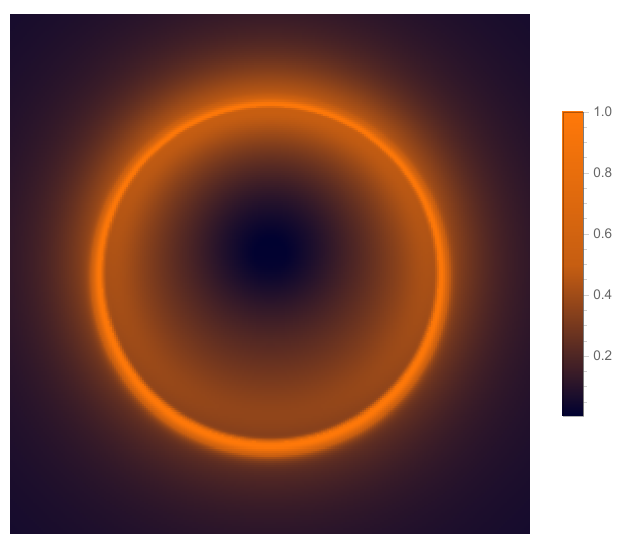}}
    
    \subfigure[$\eta=-1,\theta_o=60^\circ$]{\includegraphics[scale=0.45]{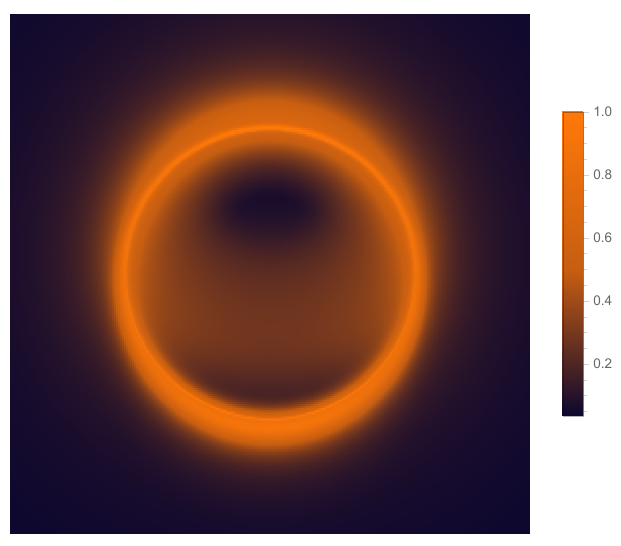}}
	\subfigure[$\eta=0.1,\theta_o=60^\circ$]{\includegraphics[scale=0.45]{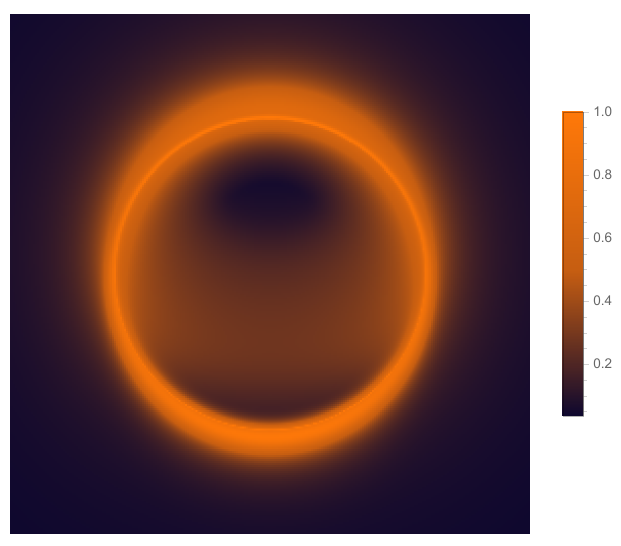}}
	\subfigure[$\eta=2,\theta_o=60^\circ$]{\includegraphics[scale=0.45]{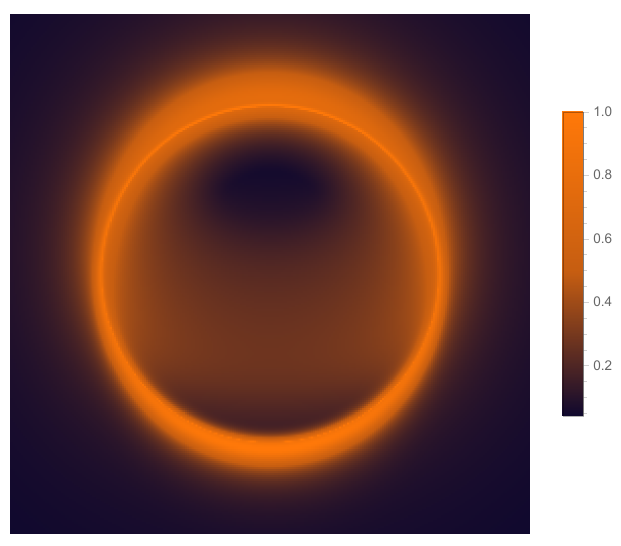}}

     \subfigure[$\eta=-1,\theta_o=85^\circ$]{\includegraphics[scale=0.45]{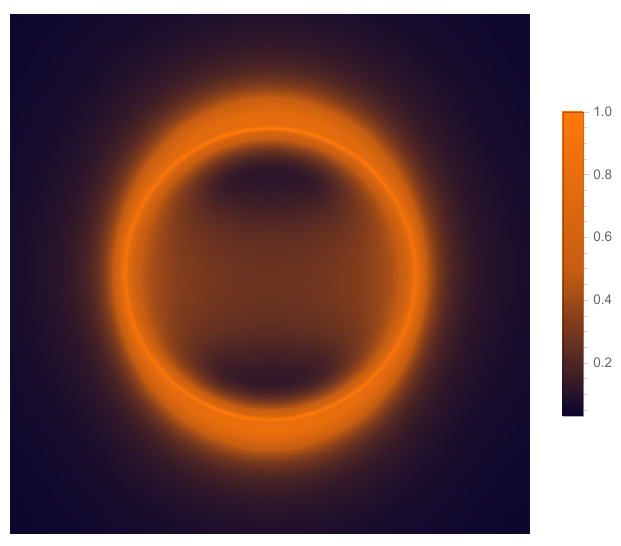}}
	\subfigure[$\eta=0.1,\theta_o=85^\circ$]{\includegraphics[scale=0.45]{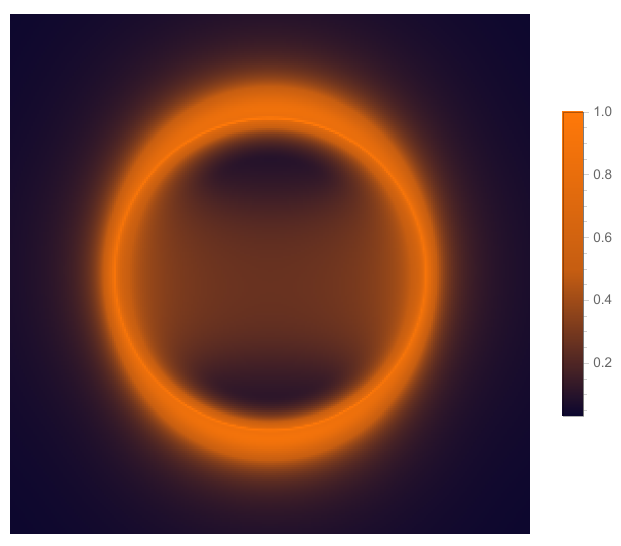}}
	\subfigure[$\eta=2,\theta_o=85^\circ$]{\includegraphics[scale=0.45]{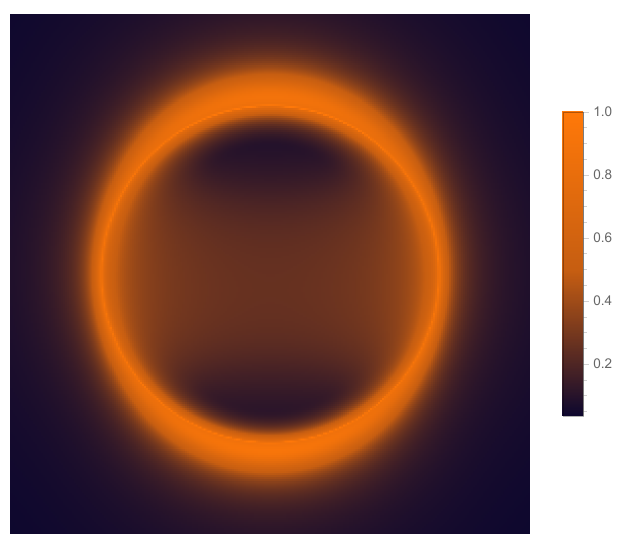}}
    \caption{Intensity maps of the KZ black hole in the RIAF model with anisotropic emission. The accretion flow follows the infalling motion, the observing frequency is 230 GHz.}
    \label{fig:7}
\end{figure}

\begin{figure}[htbp]
	\centering

    \subfigure[$\theta_o=0.001^\circ$]{\includegraphics[scale=0.85]{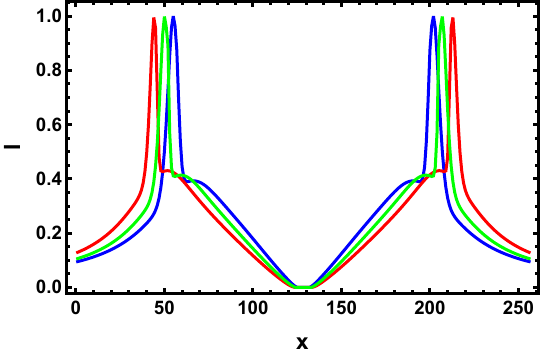}}
	\subfigure[$\theta_o=17^\circ$]{\includegraphics[scale=0.85]{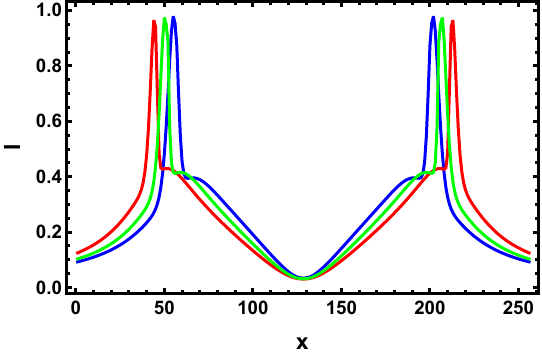}}
	\subfigure[$\theta_o=60^\circ$]{\includegraphics[scale=0.85]{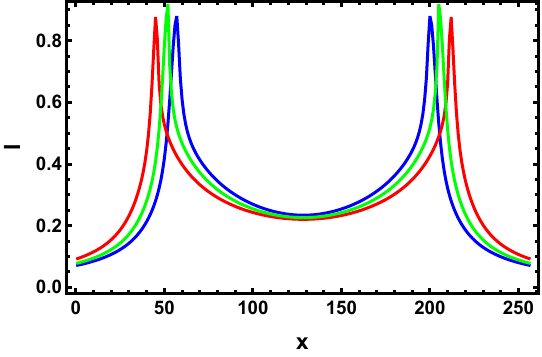}}
    \subfigure[$\theta_o=85^\circ$]{\includegraphics[scale=0.85]{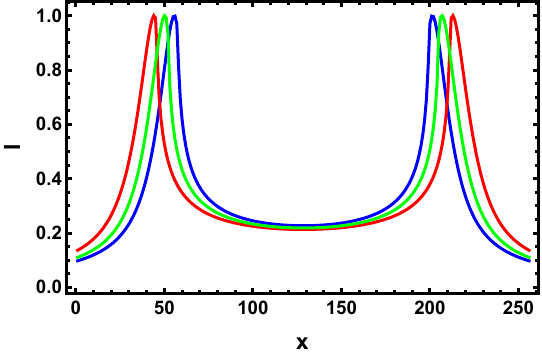}}

    \caption{Intensity distribution along the screen $x$-axis for the KZ black hole in the RIAF model with anisotropic emission. 
     The accretion flow follows the infalling motion, and the observing frequency is fixed at $230\,\mathrm{GHz}$. 
     The curves correspond to different values of $\eta$: red for $\eta=2$, green for $\eta=0.1$, and blue for $\eta=-1$.}
     \label{fig:8}
\end{figure}
\begin{figure}[htbp]
	\centering 
\subfigure[$\theta_o=0.001^\circ$]{\includegraphics[scale=0.85]{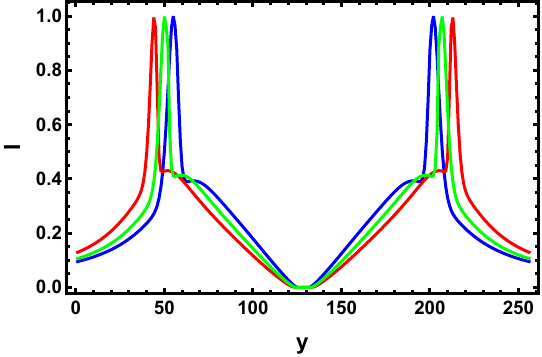}}
	\subfigure[$\theta_o=17^\circ$]{\includegraphics[scale=0.85]{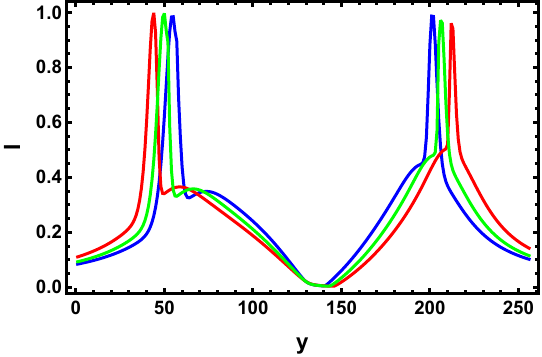}}
	\subfigure[$\theta_o=60^\circ$]{\includegraphics[scale=0.85]{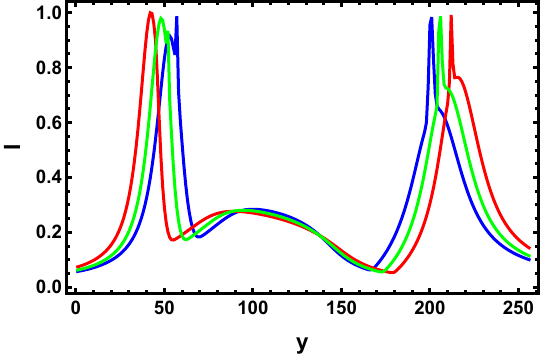}}
    \subfigure[$\theta_o=85^\circ$]{\includegraphics[scale=0.85]{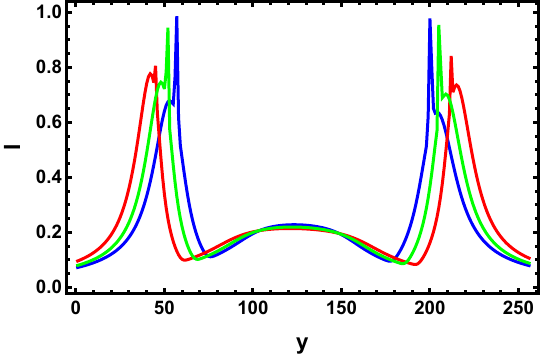}}

    \caption{Intensity distribution along the screen $y$-axis for the KZ black hole in the RIAF model with anisotropic emission. 
     The accretion flow follows the infalling motion, and the observing frequency is fixed at $230\,\mathrm{GHz}$. 
     The curves correspond to different values of $\eta$: red for $\eta=2$, green for $\eta=0.1$, and blue for $\eta=-1$.}
     \label{fig:9}
\end{figure}

Fig.~\ref{fig:7} shows the intensity maps of the KZ black hole in the RIAF model with anisotropic synchrotron emission.
The accretion flow follows the infalling motion, and the observing frequency is fixed at $230\,\mathrm{GHz}$.
Rows correspond to different inclination inclinations, $\theta_o = 0.001^\circ$, $17^\circ$, $60^\circ$, and $85^\circ$, while columns vary the deformation parameter as $\eta = -1$, $0.1$, and $2$.
To facilitate quantitative comparisons, the corresponding horizontal and vertical intensity profiles are displayed in Fig.~\ref{fig:8} and Fig.~\ref{fig:9}, respectively.
The overall morphology closely resembles that of the isotropic case (Fig.~\ref{fig:1}), featuring a pronounced bright ring encircling a dark central region—both of which expand with increasing $\eta$.
At higher inclination angles, the brightness distribution becomes strongly nonuniform, and two dark zones emerge within the ring.
A key distinction of the anisotropic case is the development of a vertically elongated and elliptical ring structure at large inclinations.
This asymmetry originates from the angular dependence of synchrotron emissivity: for photons emitted from the upper and lower regions of the disk, the propagation direction is nearly perpendicular to the magnetic field, enhancing the emission and stretching the ring vertically. It is important to note that this intensity distribution arises due to the assumption of a purely toroidal magnetic field. If poloidal components were included, the pitch angle would change, which would alter the asymmetry in the intensity distribution. As shown in Fig.~\ref{fig:15} of the Appendix~\ref{appendix:A}, the introduction of poloidal components leads to variations in the observed brightness distribution, highlighting how different magnetic field topologies influence the ring’s appearance.

\section{BAAF Model}
\label{sec5}

In the previous section, we examined the imaging characteristics of the KZ black hole surrounded by a phenomenological RIAF thick disk, considering both isotropic and anisotropic synchrotron emission. We now turn to a more analytically tractable model-the \textit{BAAF}-to further explore the geometric and radiative properties of thick accretion disk.

The BAAF disk~\cite{Hou:2023bep} describes a steady, axisymmetric accretion configuration in which the flow satisfies $u^\theta \equiv 0$, indicating that the streamlines are confined to constant-$\theta$ surfaces. Under this assumption, the mass conservation equation reduces to
\begin{equation}
\frac{\mathrm{d}}{\mathrm{~d} r}\left(\sqrt{-g} \rho u^r\right)=0, \quad \Longrightarrow \rho=\rho_0 \frac{\left.\sqrt{-g} u^r\right|_{r=r_0}}{\sqrt{-g} u^r}\,,
\end{equation}
where $\rho_0=\rho(r_0)$ is the rest mass density at a reference radius $r_0$, typically chosen to be the event horizon, $r_0=r_{h}$. Moreover, the projection of the energy–momentum conservation equation $\nabla_\mu T^{\mu\nu}=0$ along the four-velocity yields
\begin{equation}
\label{energy}
\df \Xi=\frac{\Xi+p}{\rho} \df \rho\,,
\end{equation}
where $\Xi$ denotes the internal energy density of the fluid.  
Defining the proton-to-electron temperature ratio $z = T_p/T_e$, which quantifies the relative thermal states of the two species in the plasma, the internal energy density satisfies
\begin{equation}
\label{xi}
\Xi=\rho+\rho \frac{3}{2}(z+2) \frac{m_e}{m_p} \theta_e\,,
\end{equation}
where, as before, $\theta_e$ denotes the dimensionless electron temperature.  
Using the ideal gas equation of state, one obtains
\begin{equation}
\label{gas}
p=n k_B\left(T_p+T_e\right)=\rho(1+z) \frac{m_e}{m_p} \theta_e\,,
\end{equation}
and substituting Eqs.~\eqref{xi} and~\eqref{gas} into Eq.~\eqref{energy}, one obtains after integration
\begin{equation}
\theta_e=\left(\theta_e\right)_0\left(\frac{\rho}{\rho_0}\right)^{\frac{2(1+z)}{3(2+z)}}\,,
\end{equation}
where $(\theta_e)_0=\theta_e(r_0)$.

Assuming an infalling flow that satisfies Eq.~\eqref{infalling}, the analytical expressions for the rest-mass density and the electron temperature become
\begin{equation}
\begin{aligned}
    \rho(r,\theta) &=\rho(r_h,\theta) \left(\frac{r_h}{r}\right)^2 \sqrt{\frac{1}{1-g^{rr}}}=\rho(r_h,\theta) \sqrt{\frac{r_h^4}{2 r^3 +\eta r}}\,,\\
    \theta_e(r,\theta) &=\theta_e(r_h,\theta)\left(\frac{r_h}{r}\right)^{\frac{4(1+z)}{3(2+z)}} \left(\frac{1}{1-g^{rr}}\right)^{\frac{1+z}{3(2+z)}}=\theta_e(r_h,\theta) \left( \frac{r_h^4}{2 r^3 +\eta r}\right)^{\frac{1+z}{3(2+z)}}\,.
\end{aligned}
\end{equation}

For convenience in subsequent analysis, the angular dependence of $\rho(r_h,\theta)$ is modeled by a Gaussian profile, while in the conical solution we take $\theta_e(r_h,\theta)$ to be constant:
\begin{equation}
\rho\left(r_h, \theta\right)=\rho_h \exp \left[-\left(\frac{\sin \theta-\sin \theta_J}{\sigma}\right)^2\right], \quad \theta\left(r_h, \theta\right)=\theta_h\,.
\end{equation}
Here, $\theta_J$ specifies the central latitude of the distribution and $\sigma$ its angular width.  
We consider an equatorially symmetric thick disk with $\theta_J=\pi/2$ and $\sigma=1/5$. For the $\text{M87}^*$ black hole, observational estimates suggest
$\rho_h \simeq 1.5 \times 10^3~\mathrm{g\,cm^{-3}\,s^{-2}}$, and $\theta_h \simeq 16.86$,
corresponding to 
$n_h \simeq 10^6~\mathrm{cm^{-3}}$, $T_h \simeq 10^{11}~\mathrm{K}$.

Assuming stationarity, axisymmetry, and the ideal MHD condition~\cite{Ruffini:1975ne}, the magnetic field takes the general form~\cite{Ruffini:1975ne,Hou:2023bep}
\begin{equation}
\label{beq}
B^\mu=\frac{\Psi}{\sqrt{-g} u^r}\left(\left(u_t+\Omega_B u_\phi\right) u^\mu+\delta_t^\mu+\Omega_B \delta_\phi^\mu\right)\,,
\end{equation}
where $\Psi=F_{\theta\phi}$ is a component of the electromagnetic tensor and $\Omega_B$ denotes the field line angular velocity, which characterizes the rotation  of magnetic field lines. It is evident that the spatial components $B^i$ are parallel to $u^i$
along the poloidal direction, indicating that the magnetic field is effectively frozen into the streamlines. For simplicity and without loss of generality, we set $\Omega_B=0$ in what follows.  
In this work, we adopt a separable magnetic monopole configuration, which is the simplest model for the magnetic field configuration~\cite{Blandford:1977ds}
\begin{equation}
\Psi=\Psi_0 \operatorname{sign}(\cos \theta) \sin \theta\,,
\end{equation}
where $\Phi_0$ is a constant. The inclusion of the factor $\sin\theta$ ensures regularity at the poles where $\sin\theta=0$. The $\operatorname{sign}(\cos \theta)$ guarantees that the black hole carries zero net magnetic charge. Numerical studies have shown that such a magnetic field structure naturally emerges in the near-horizon region with a initial uniform  magnetic field~\cite{Komissarov:2004qu,Komissarov:2004ms}.

Following the characteristic field strength at $(r_h, \theta_J)$ used in~\cite{Zhang:2024lsf}, we take $B_i = 10~\text{Gauss}$. For simplicity, we set $z=20$~\cite{Zhang:2024lsf} in the subsequent analysis, corresponding to a regime where the thermal motion of protons remains non-relativistic. Following~\cite{Moscibrodzka:2015pda}, a commonly used GRMHD prescription for the proton-to-electron temperature ratio is $T_{\rm p}/T_{\rm e} = R_{\rm high}\,\beta^2/(1+\beta^2) + 1/(1+\beta^2)$, where $\beta \equiv p_{\rm gas}/p_M$ and $R_{\rm high}$ is a phenomenological parameter typically ranging from 1 to 160 for M87*~\cite{EventHorizonTelescope:2019pgp}. Using $p_M \approx B_i^2/2 \approx 0.72\times 10^{52}$ and $p_{\rm gas} \approx n_i k_B T_{\rm e} (1+z) \approx 10^{51} (1+z)$~\cite{Zhang:2024lsf}, one finds $\beta^2 \gg 1$. In this weakly magnetized regime, the temperature ratio reduces to $T_{\rm p}/T_{\rm e} \approx R_{\rm high}$, justifying the characteristic temperature ratio adopted in our analysis. We note that although $z$ may vary
spatially and span a wide range for M87*, the polarization
vectors are predominantly oriented perpendicular to both the local magnetic
field and the radiation propagation direction.
As a result, the polarization patterns are primarily governed by
the spacetime geometry and the magnetic field configuration, while being only
weakly sensitive to the specific value of $z$.

Finally, based on the methodology introduced earlier, we present the black hole images illuminated by the BAAF disk and analyze their corresponding intensity and polarization distributions.
\begin{figure}[htbp]
	\centering 
	\subfigure[$\eta=-1,\theta_o=0.001^\circ$]{\includegraphics[scale=0.45]{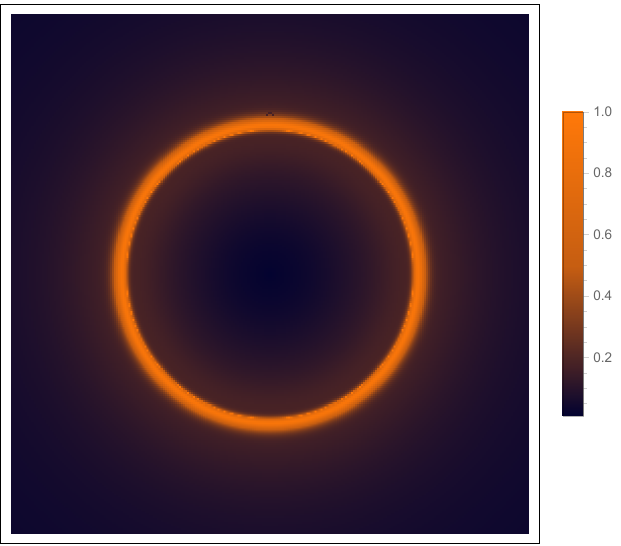}}
	\subfigure[$\eta=0.1,\theta_o=0.001^\circ$]{\includegraphics[scale=0.45]{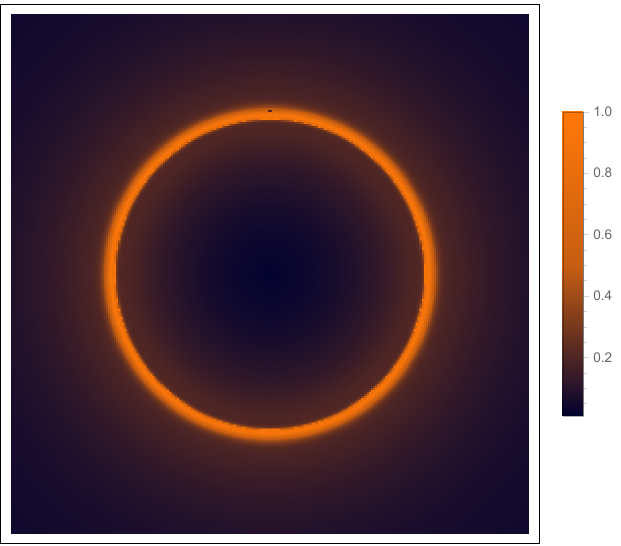}}
	\subfigure[$\eta=2,\theta_o=0.001^\circ$]{\includegraphics[scale=0.45]{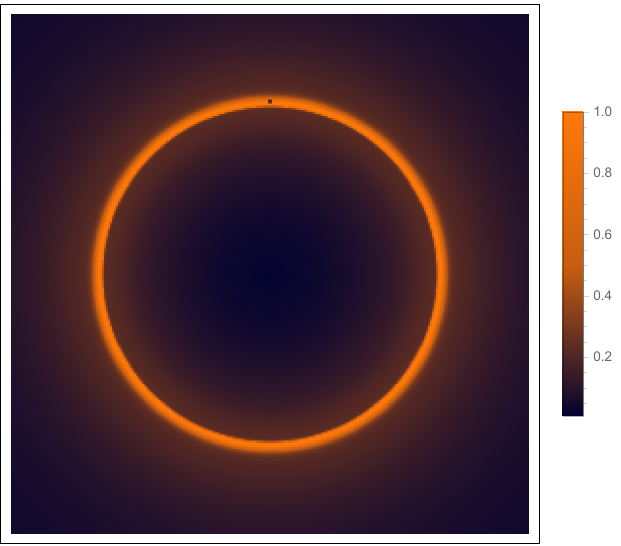}}

\subfigure[$\eta=-1,\theta_o=17^\circ$]{\includegraphics[scale=0.45]{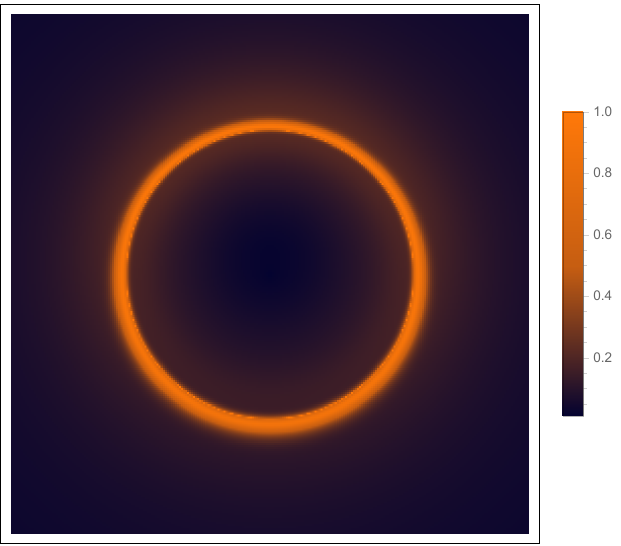}}
	\subfigure[$\eta=0.1,\theta_o=17^\circ$]{\includegraphics[scale=0.45]{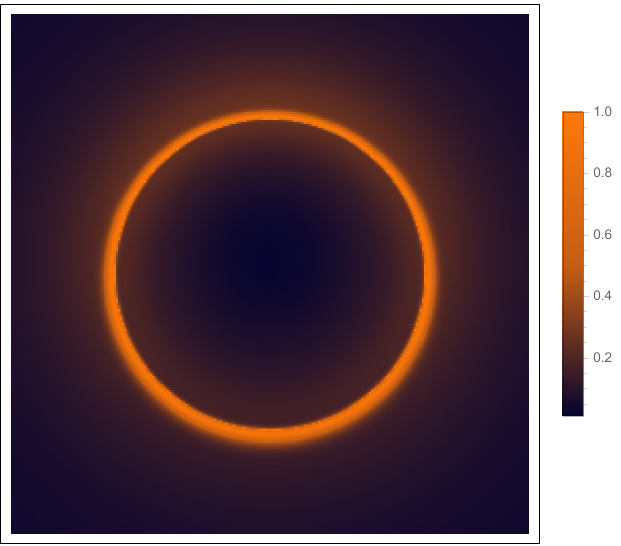}}
	\subfigure[$\eta=2,\theta_o=17^\circ$]{\includegraphics[scale=0.45]{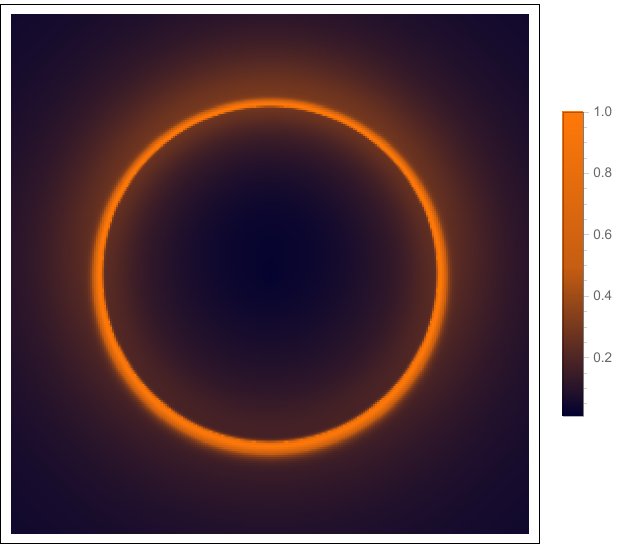}}
    
    \subfigure[$\eta=-1,\theta_o=60^\circ$]{\includegraphics[scale=0.45]{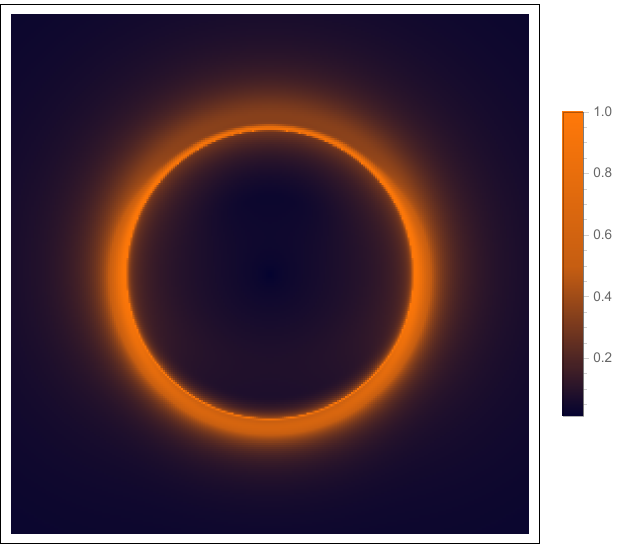}}
	\subfigure[$\eta=0.1,\theta_o=60^\circ$]{\includegraphics[scale=0.45]{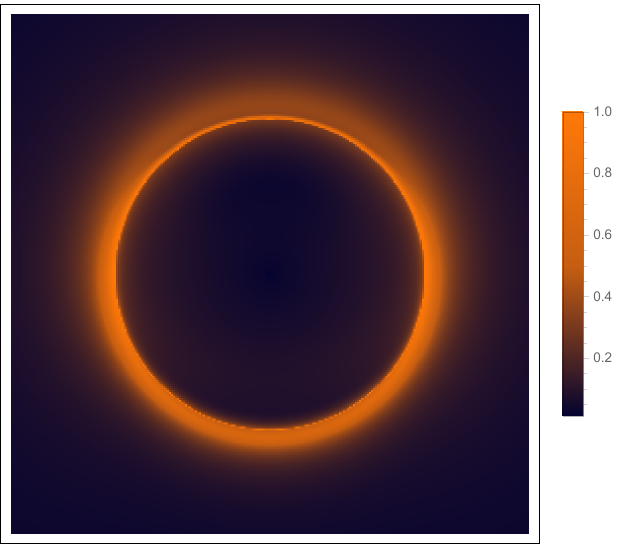}}
	\subfigure[$\eta=2,\theta_o=60^\circ$]{\includegraphics[scale=0.45]{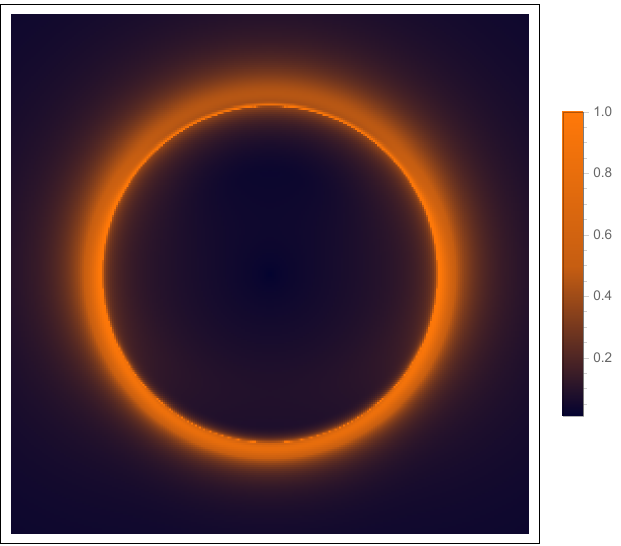}}

     \subfigure[$\eta=-1,\theta_o=85^\circ$]{\includegraphics[scale=0.45]{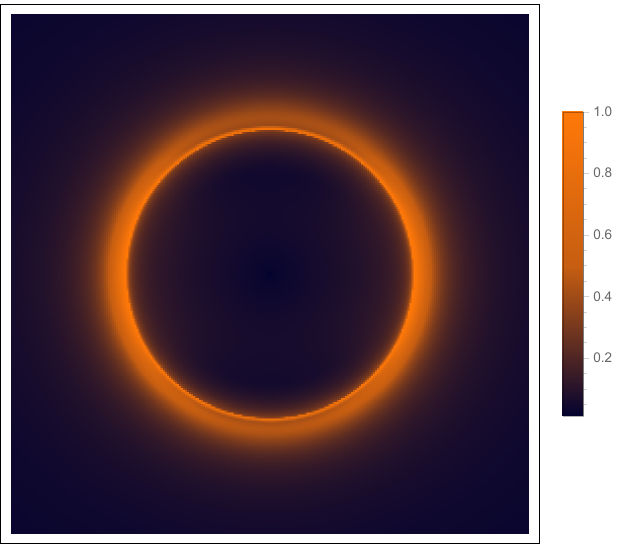}}
	\subfigure[$\eta=0.1,\theta_o=85^\circ$]{\includegraphics[scale=0.45]{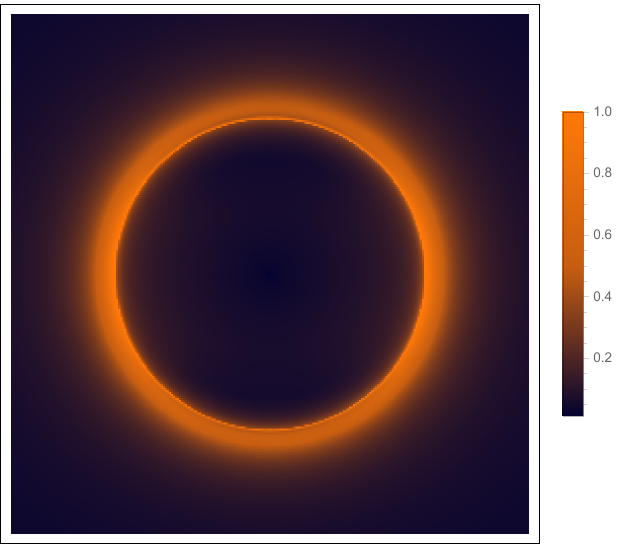}}
	\subfigure[$\eta=2,\theta_o=85^\circ$]{\includegraphics[scale=0.45]{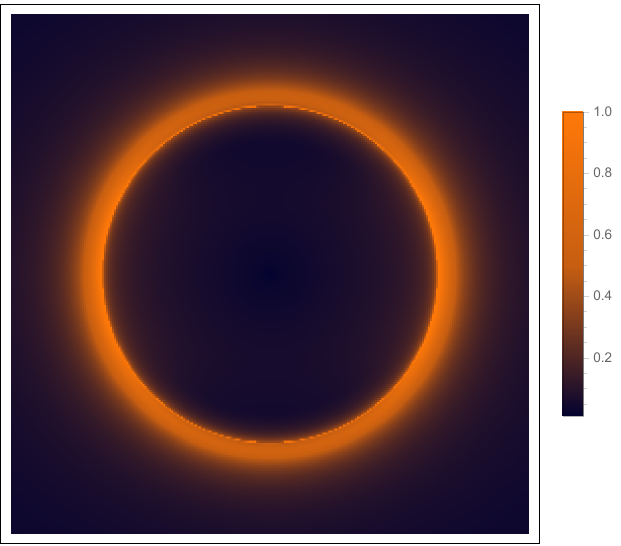}}
    \caption{Intensity maps of the KZ black hole in the BAAF model with anisotropic emission. The accretion flow follows the infalling motion, the observing frequency is 230 GHz.}
    \label{fig:10}
\end{figure}

\begin{figure}[htbp]
	\centering

    \subfigure[$\theta_o=0.001^\circ$]{\includegraphics[scale=0.85]{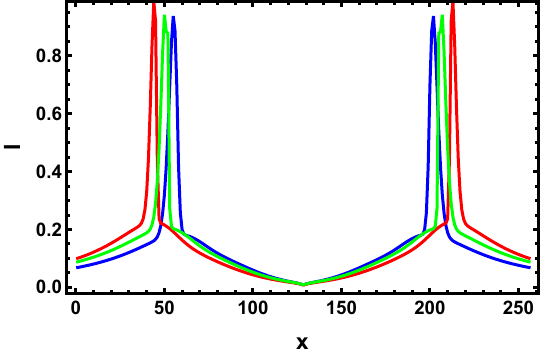}}
	\subfigure[$\theta_o=17^\circ$]{\includegraphics[scale=0.85]{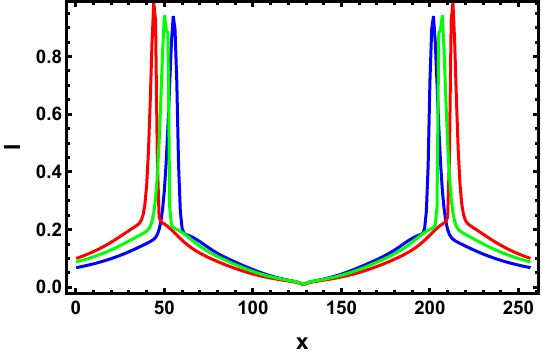}}
	\subfigure[$\theta_o=60^\circ$]{\includegraphics[scale=0.85]{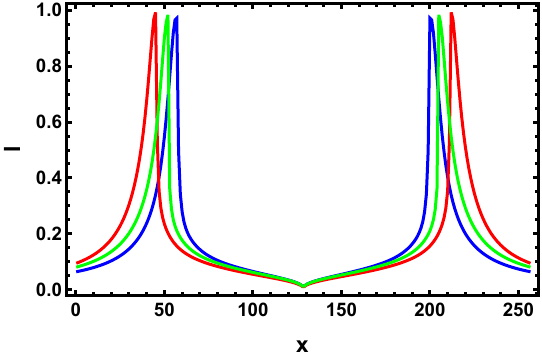}}
    \subfigure[$\theta_o=85^\circ$]{\includegraphics[scale=0.85]{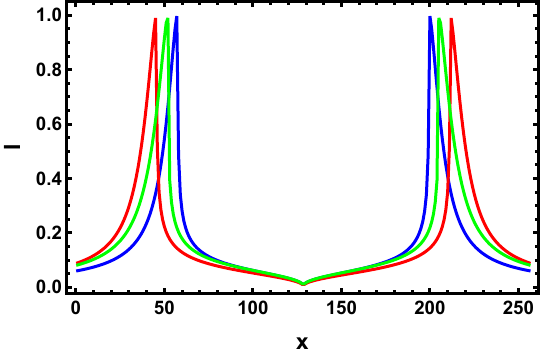}}

    \caption{Intensity distribution along the screen $x$-axis for the KZ black hole in the BAAF model with anisotropic emission. 
     The accretion flow follows the infalling motion, and the observing frequency is fixed at $230\,\mathrm{GHz}$. 
     The curves correspond to different values of $\eta$: red for $\eta=2$, green for $\eta=0.1$, and blue for $\eta=-1$.}
     \label{fig:11}
\end{figure}

\begin{figure}[htbp]
	\centering 
\subfigure[$\theta_o=0.001^\circ$]{\includegraphics[scale=0.85]{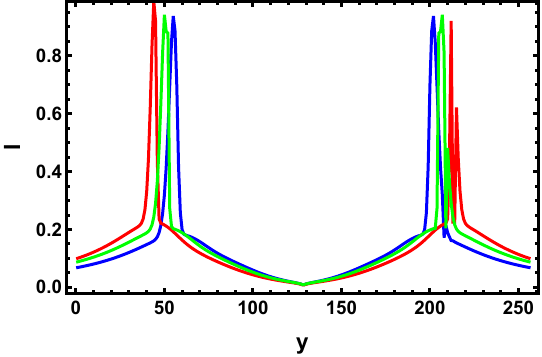}}
	\subfigure[$\theta_o=17^\circ$]{\includegraphics[scale=0.85]{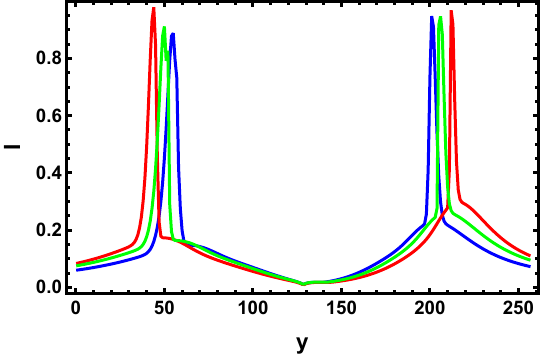}}
	\subfigure[$\theta_o=60^\circ$]{\includegraphics[scale=0.85]{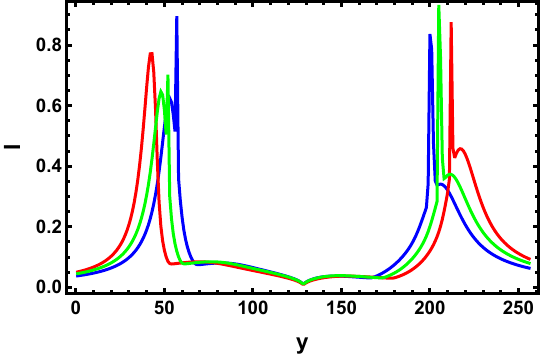}}
    \subfigure[$\theta_o=85^\circ$]{\includegraphics[scale=0.85]{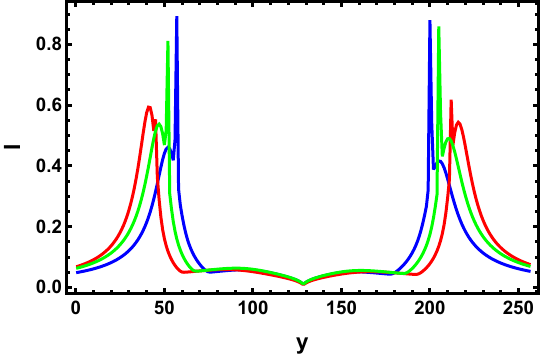}}

    \caption{Intensity distribution along the screen $y$-axis for the KZ black hole in the BAAF model with anisotropic emission. 
     The accretion flow follows the infalling motion, and the observing frequency is fixed at $230\,\mathrm{GHz}$. 
     The curves correspond to different values of $\eta$: red for $\eta=2$, green for $\eta=0.1$, and blue for $\eta=-1$.}
    \label{fig:12}
\end{figure}
Fig.~\ref{fig:10} illustrates the total intensity maps of the KZ black hole obtained from the BAAF disk model with anisotropic synchrotron emission. The accretion flow is assumed to follow a purely infalling motion, and the observing frequency is fixed at $230\,\mathrm{GHz}$. 
Each panel corresponds to a specific combination of the deformation parameter $\eta$ and the observer’s inclination angle $\theta_o$, as indicated. 
The corresponding horizontal and vertical intensity profiles are presented in Figs.~\ref{fig:11} and~\ref{fig:12}, respectively, providing a quantitative representation of the brightness distribution.
The bright ring visible in all images originates from higher-order images, while the central dark region corresponds to the event horizon. 
Overall, the dependence of image morphology on the deformation parameter $\eta$ and inclination angle $\theta_o$ closely resembles that observed in the RIAF thick disk model: 
increasing $\eta$ enlarges both the bright ring and the inner dark region. 
The dependence on the inclination angle likewise follows the symmetry trends discussed previously. 
However, a few notable distinctions emerge. 
Compared with the RIAF model, the bright ring in the BAAF disk case appears generally thinner, and the separation between the primary and higher-order images becomes more pronounced. 
Furthermore, at large inclination angles, the higher-order image does not exhibit the two distinct dark zones observed in the RIAF case. 
This suggests that, in the RIAF model, radiation from off-equatorial regions more effectively obscures the event-horizon silhouette. 
Such differences may stem from the fact that, for certain parameter choices, the BAAF disk in the conical approximation is geometrically thinner than the RIAF disk in some regions.

\begin{figure}[htbp]
	\centering 
\subfigure[$Stokes$ $\mathcal{I}_o$]{\includegraphics[scale=0.48]{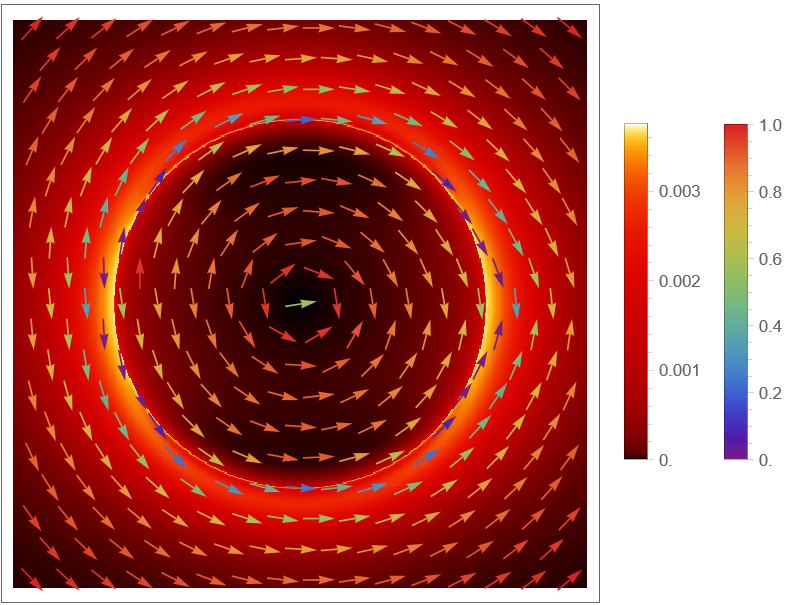}}
	\subfigure[$Stokes$ $\mathcal{Q}_o$]{\includegraphics[scale=0.5]{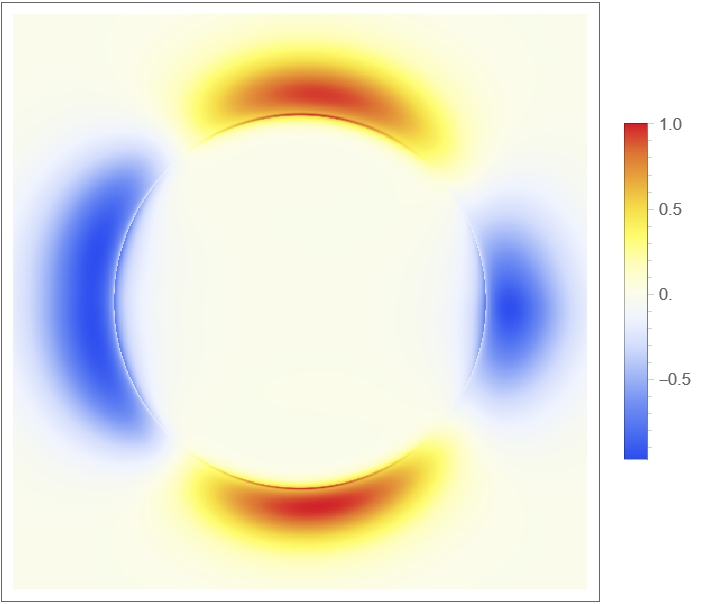}}
	\subfigure[$Stokes$ $\mathcal{U}_o$]{\includegraphics[scale=0.5]{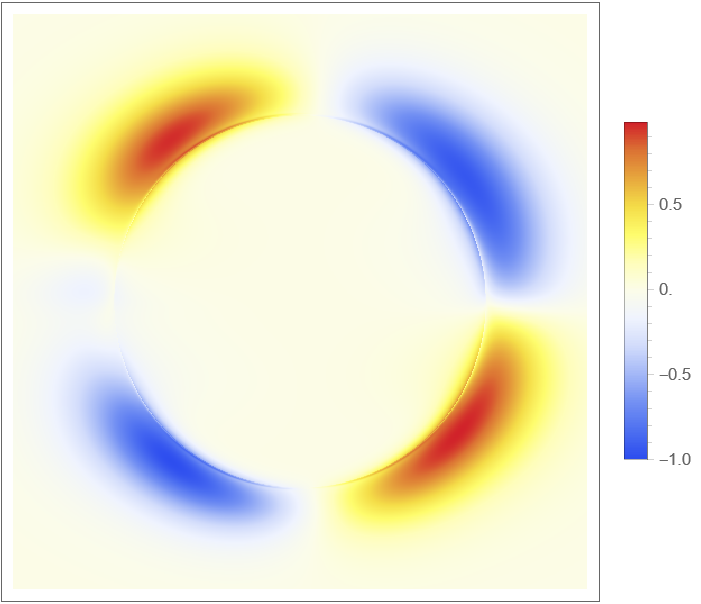}}
    \subfigure[$Stokes$ $\mathcal{V}_o$]{\includegraphics[scale=0.5]{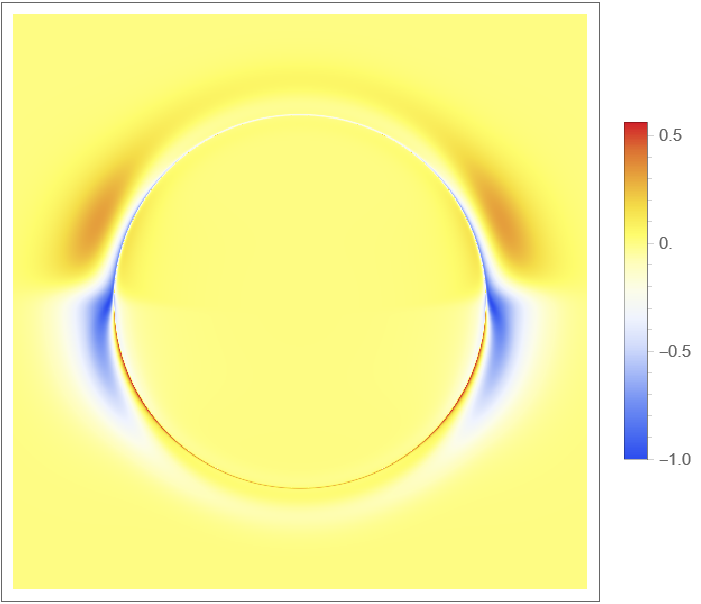}}

    \caption{The resulting Stokes parameters $\mathcal{I}_o$, $\mathcal{Q}_o$, $\mathcal{U}_o$, $\mathcal{V}_o$ under the BAAF disk model. The dynamics of the accretion flow is infalling motion, with fixed parameters $\eta=2\,, \theta_o= 85^\circ\,,230\, \mathrm{GHz}$.}
\label{fig:13}
\end{figure}
We now turn to the analysis of the polarization properties. 
Fig.~\ref{fig:13} presents the spatial distributions of the observed Stokes parameters, 
$\mathcal{I}_o$, $\mathcal{Q}_o$, $\mathcal{U}_o$, and $\mathcal{V}_o$, 
for the KZ black hole in the BAAF model. 
The accretion flow is assumed to be purely infalling, 
with fixed parameters $\eta = 2$, $\theta_o = 85^\circ$, 
and an observing frequency of $230\,\mathrm{GHz}$. 
The $\mathcal{I}_o$ panel shows the total intensity distribution, 
where the arrows indicate the electric vector position angle (EVPA), 
$\Phi_{\mathrm{EVPA}}$, and the color encodes the linear polarization degree, $P_o$. Since the EVPA is predominantly perpendicular to the magnetic field, $B^\mu$, 
the polarization pattern suggests that the magnetic field is roughly radial. 
The combined distributions of $\mathcal{Q}_o$ and $\mathcal{U}_o$ qualitatively determine 
the EVPA orientation, while $\mathcal{V}_o < 0$ corresponds to right-handed circular polarization.

\begin{figure}[htbp]
	\centering

    \subfigure[$\eta=-1,\theta_o=0.001^\circ$]{\includegraphics[scale=0.3]{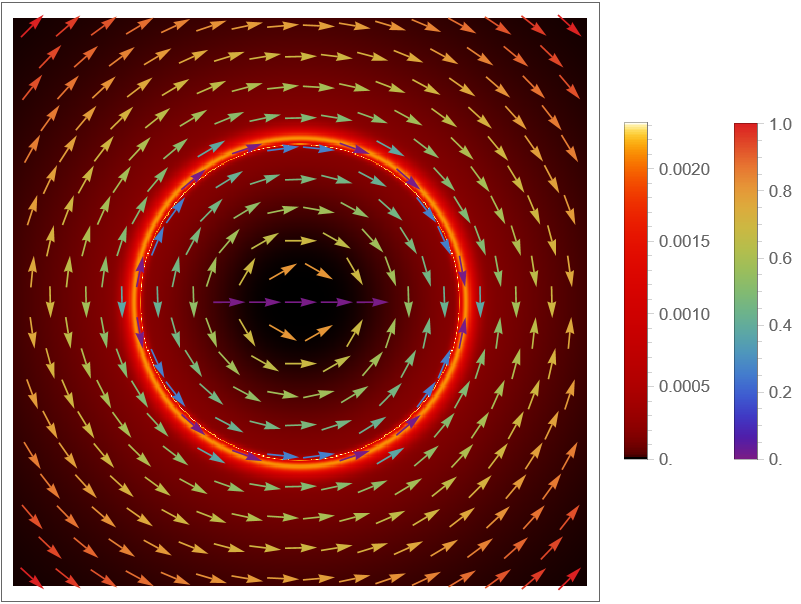}}
	\subfigure[$\eta=0.1,\theta_o=0.001^\circ$]{\includegraphics[scale=0.3]{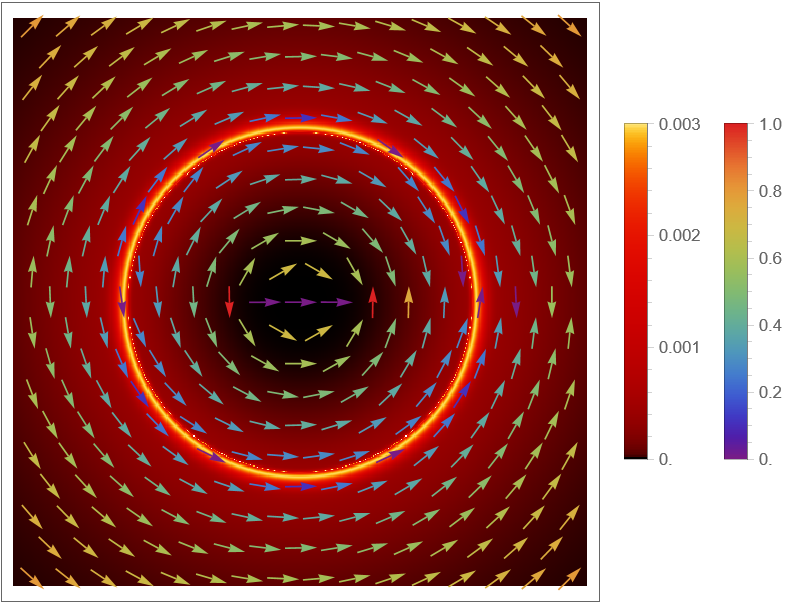}}
	\subfigure[$\eta=2,\theta_o=0.001^\circ$]{\includegraphics[scale=0.3]{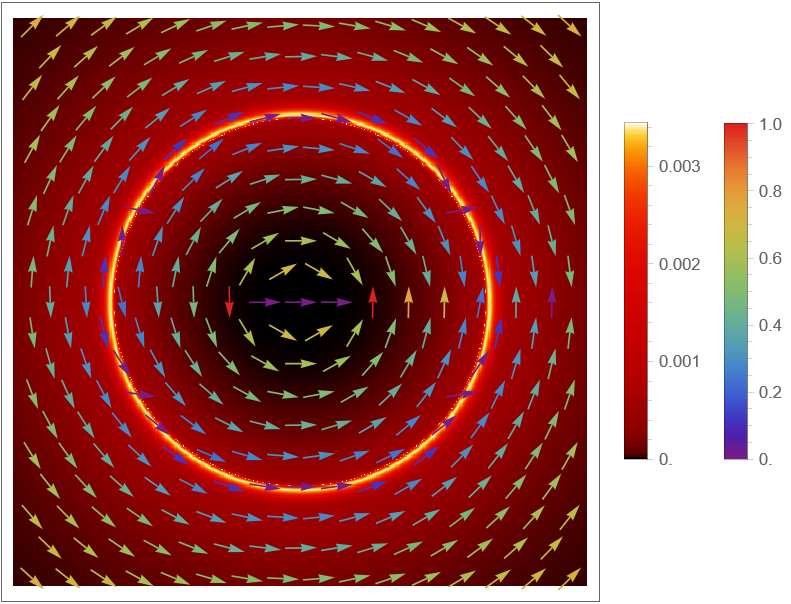}}

\subfigure[$\eta=-1,\theta_o=17^\circ$]{\includegraphics[scale=0.3]{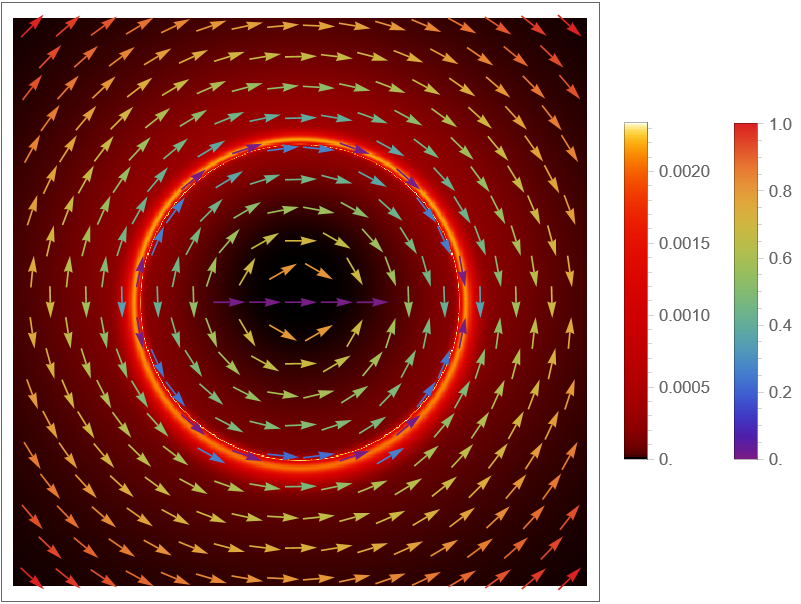}}
	\subfigure[$\eta=0.1,\theta_o=17^\circ$]{\includegraphics[scale=0.3]{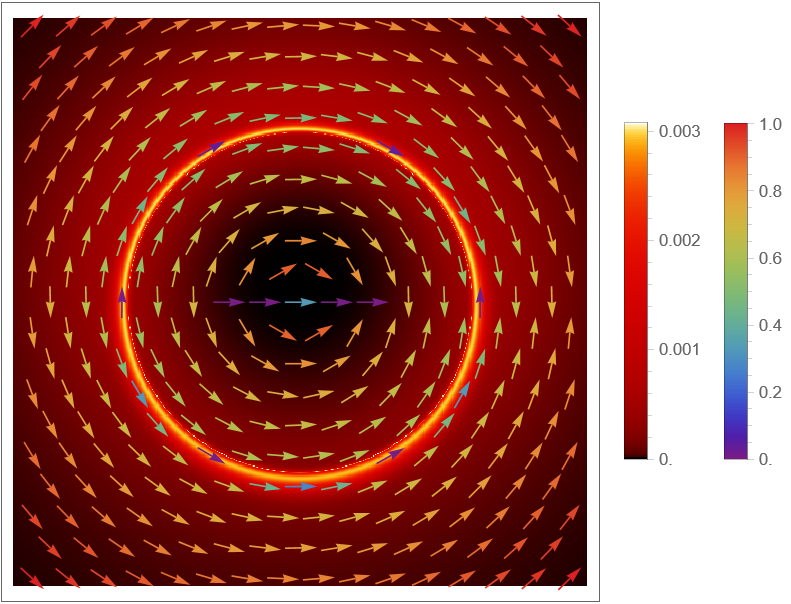}}
	\subfigure[$\eta=2,\theta_o=17^\circ$]{\includegraphics[scale=0.3]{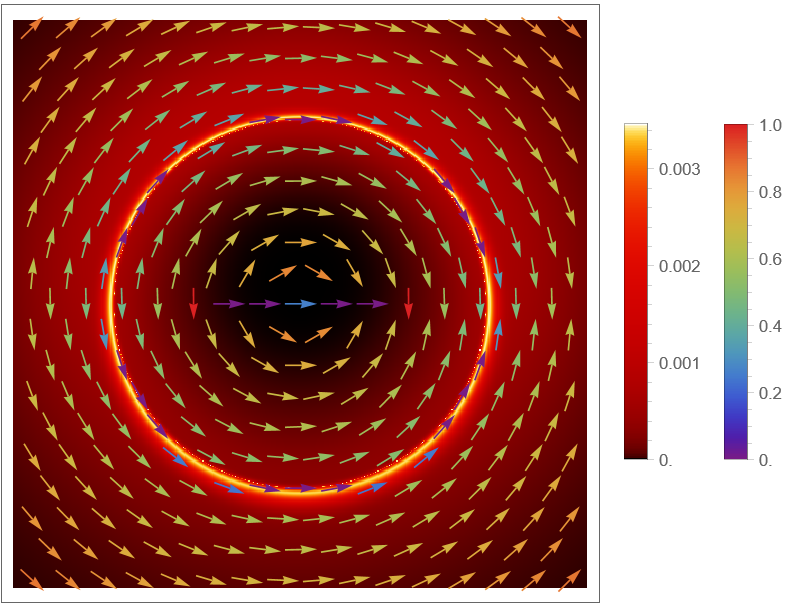}}
    
    \subfigure[$\eta=-1,\theta_o=60^\circ$]{\includegraphics[scale=0.3]{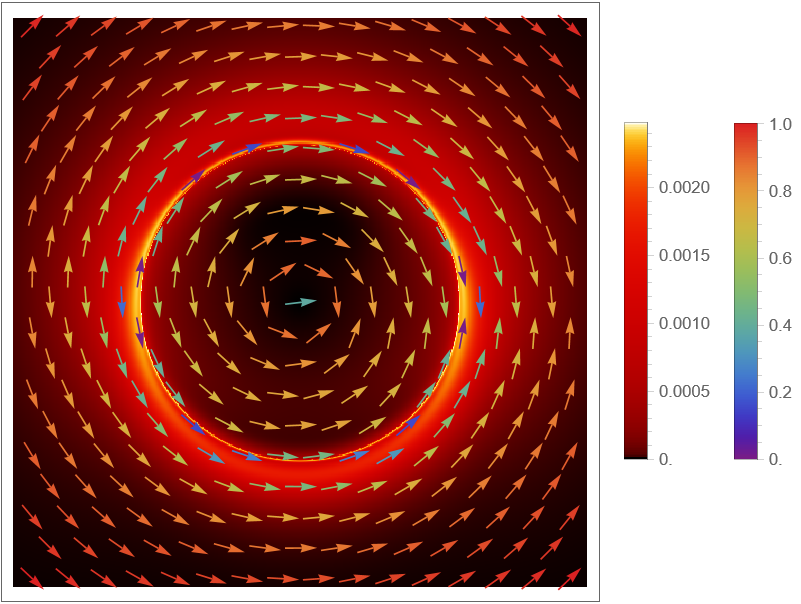}}
	\subfigure[$\eta=0.1,\theta_o=60^\circ$]{\includegraphics[scale=0.3]{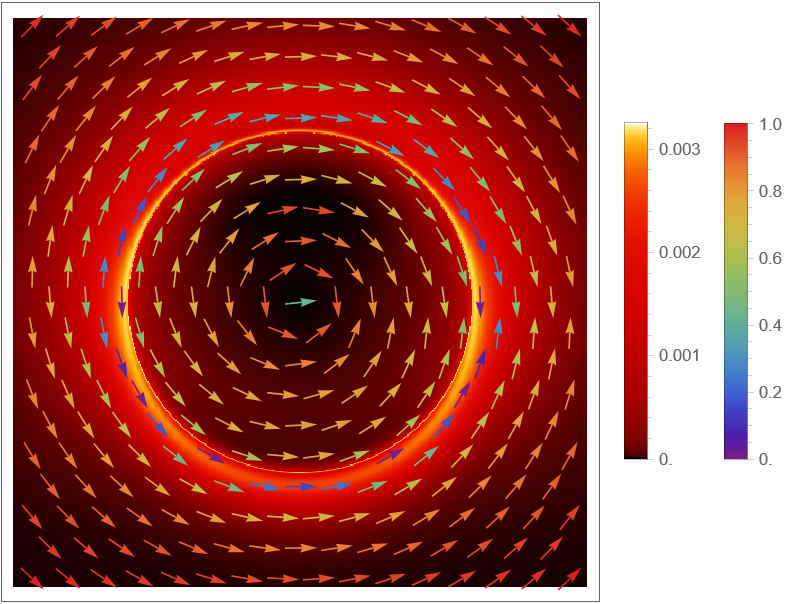}}
	\subfigure[$\eta=2,\theta_o=60^\circ$]{\includegraphics[scale=0.3]{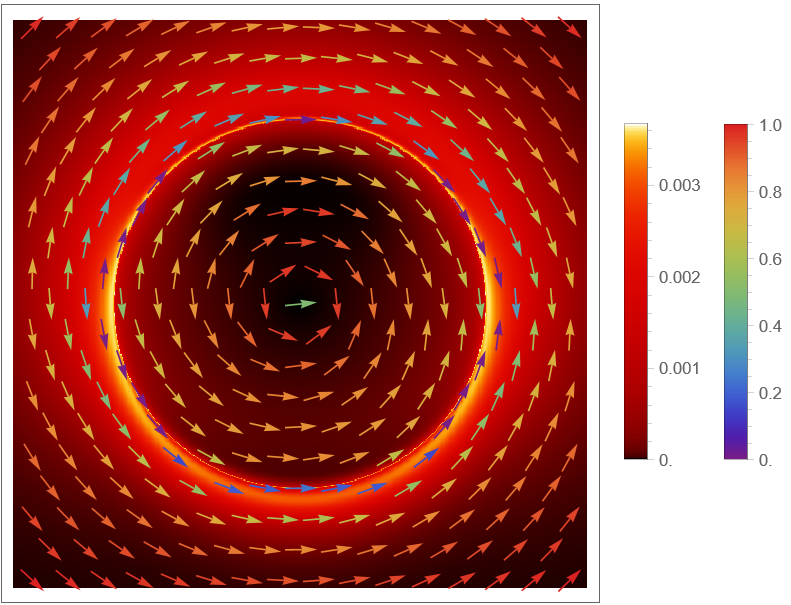}}

     \subfigure[$\eta=-1,\theta_o=85^\circ$]{\includegraphics[scale=0.3]{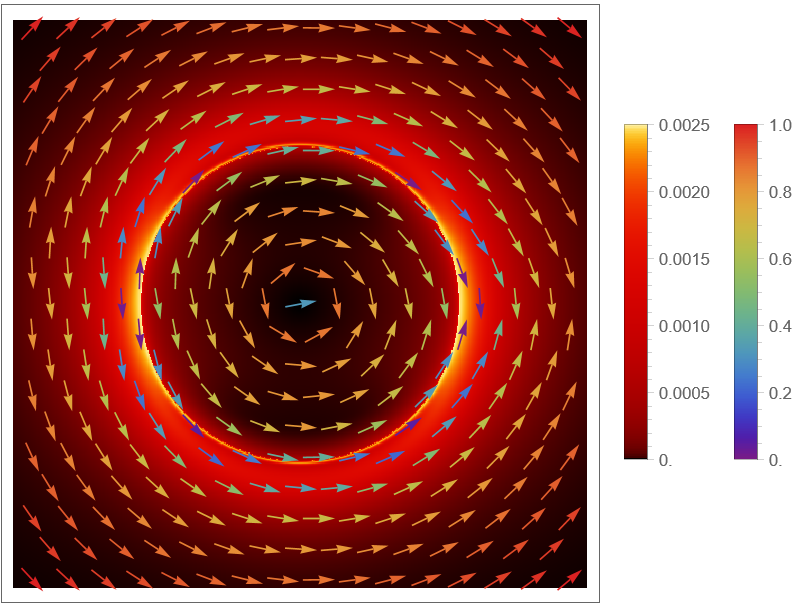}}
	\subfigure[$\eta=0.1,\theta_o=85^\circ$]{\includegraphics[scale=0.3]{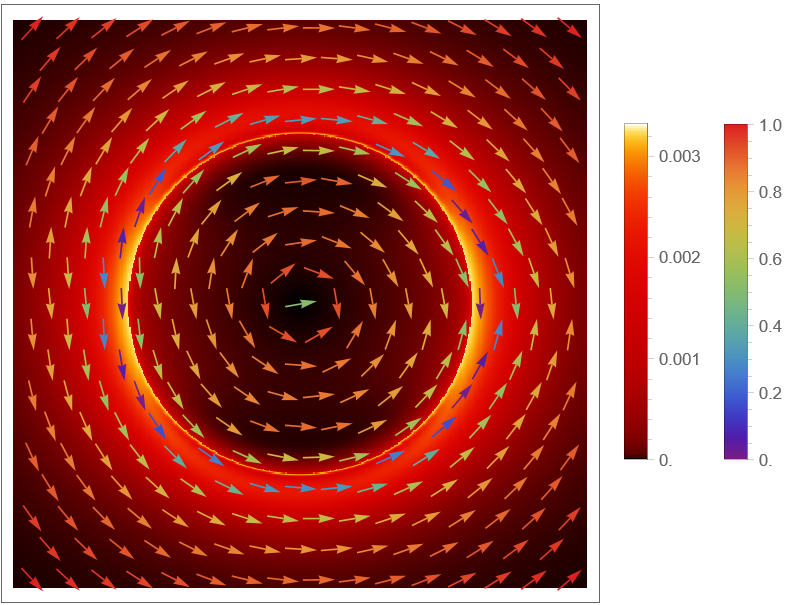}}
	\subfigure[$\eta=2,\theta_o=85^\circ$]{\includegraphics[scale=0.3]{theta=85du,eta=2,ll.png}}
    \caption{Polarized images of the KZ black hole in the BAAF model with anisotropic emission. The accretion flow follows the infalling motion, the observing frequency is 230 GHz.}
    \label{fig:14}
\end{figure}
Fig.~\ref{fig:14} presents the corresponding polarized images of the KZ black hole in the BAAF model with anisotropic emission, for various values of~$\eta$ and~$\theta_o$. The accretion flow is assumed to be purely infalling, and the images are computed at an observing frequency of $230\,\mathrm{GHz}$. The polarization patterns exhibit a clear dependence on both parameters, with significant variations in the structure and intensity of the polarized flux across the different panels. Overall, the EVPA pattern is nearly perpendicular to the radial direction, reflecting the assumed alignment of the magnetic field with the radially infalling flow. In the vicinity of the higher-order images, the polarization exhibits rapid spatial variations. Along each row of panels, as $\eta$ increases, both the size of the higher-order images and the dark region expand, accompanied by a corresponding change in the EVPA distribution. Across each column, increasing the inclination angle causes the EVPA to deviate slightly from the purely azimuthal orientation, while the polarized intensity within the dark region becomes stronger. It is noteworthy that, in thin-disk models, the dark region corresponds to the black hole horizon, from which no light reaches the observer, resulting in an absence of polarization within this region. In contrast, for the present geometrically thick disk, gravitational lensing enables emission from regions above and below the equatorial plane to obscure the horizon contour, producing polarization vectors distributed over the entire image plane.

\section{Summary}\label{sec6}

In this work, we investigated the imaging properties of spherically symmetric black holes in the Konoplya–Zhidenko (KZ) spacetime under the presence of geometrically thick accretion flows. We first reviewed the basic properties of the KZ black hole, including the numerical determination of the event horizon and the dependence of the photon sphere on the deformation parameter~$\eta$. In the small-deformation regime, we derived an approximate analytic expression for the photon ring.

We then considered two representative models of thick accretion flows: a phenomenological RIAF model and the analytical BAAF model. The synchrotron radiation from thermal electrons in the magnetofluid was computed by numerically integrating the null geodesics and radiative transfer equations to obtain the corresponding black hole images.

For the RIAF model, we first analyzed the isotropic emission case with different fluid motions—namely, orbiting, infalling, and combined motions—and several observing frequencies. The resulting images show that, as the deformation parameter~$\eta$ increases, both the photon ring and the inner dark region expand in size. For polar observers, the brightness distribution is nearly axisymmetric, whereas for larger inclination angles, the brightness becomes asymmetric and the central dark region splits into two parts due to off-equatorial emission. The morphology and contrast of the image are also sensitive to the flow dynamics and observing frequency: orbiting flows tend to blur the photon ring, making higher-order and primary images less distinguishable, while changes in observing frequency can either enhance or suppress the image contrast depending on the model parameters.

Next, we examined the case of anisotropic synchrotron emission in the RIAF model with a toroidal magnetic field and infalling flow. Compared to the isotropic case, the brightness distribution becomes noticeably asymmetric for high-inclination observers. The bright ring appears vertically elongated and slightly elliptical, reflecting the geometry of the underlying magnetic field configuration.

Finally, we analyzed the imaging and polarization properties of the BAAF disk, which assumes an infalling steady-state flow with~$u_\theta \equiv 0$. The resulting images show a narrower bright ring and a darker central region compared to the RIAF model. This difference arises because, under certain parameter choices, the BAAF disk in the conical approximation is geometrically thinner than the RIAF disk in some regions. For the polarized images, the polarization intensity closely follows the total brightness distribution, with stronger polarization in brighter regions. Both the polarization degree and orientation exhibit clear dependence on the deformation parameter and viewing angle, suggesting that the polarization signatures of KZ black holes can effectively probe the intrinsic spacetime structure. It is also worth noting that, unlike thin-disk models where the inner shadow remains unpolarized, in thick disks the gravitational lensing of off-equatorial emission allows polarization vectors to appear across the entire image plane.

In this work, we have focused on the non-rotating KZ spacetime, which provides a clean and controlled framework to isolate the effects of spacetime deformations on black hole images and polarization signatures in various thick disk models. While rotation is expected to play an important role for realistic astrophysical black holes and would be more relevant for comparisons with M87* and Sgr~A* observations, a systematic extension to the rotating KZ spacetime is left for future work.

\section*{Acknowledgments}
We are grateful to Yehui Hou, Zhenyu Zhang, Jiewei Huang, Chenyu Yang for insightful discussions. This work is supported by the National Natural Science Foundation of China (Grants Nos. 12375043,
12575069,12275004 and 12205013).

\appendix
\section{Supplementary Plots}
\label{appendix:A}
\begin{figure}[ht]
	\centering 
	\subfigure[$b^r$ only]{\includegraphics[scale=0.45]{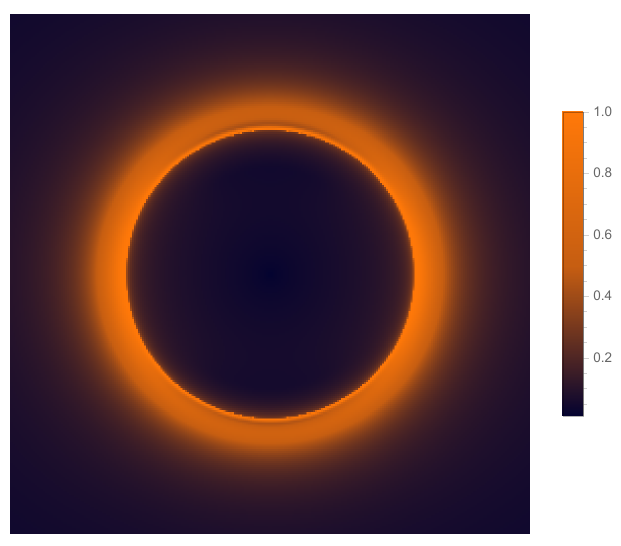}}
	\subfigure[$b^\theta$ only]{\includegraphics[scale=0.45]{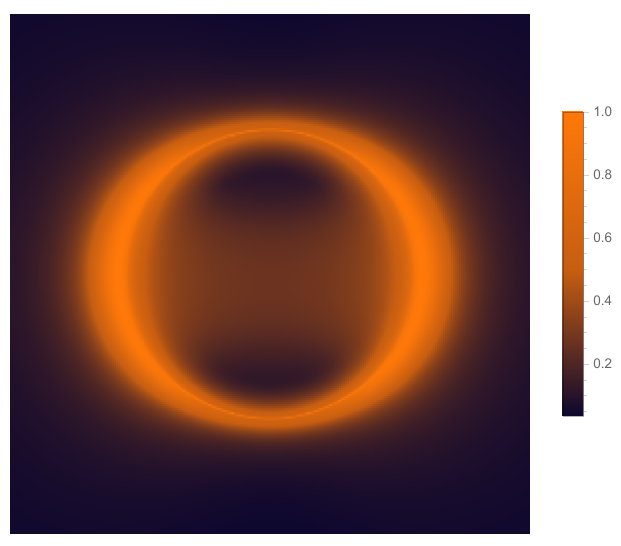}}
	\subfigure[$b^\phi$ only]{\includegraphics[scale=0.45]{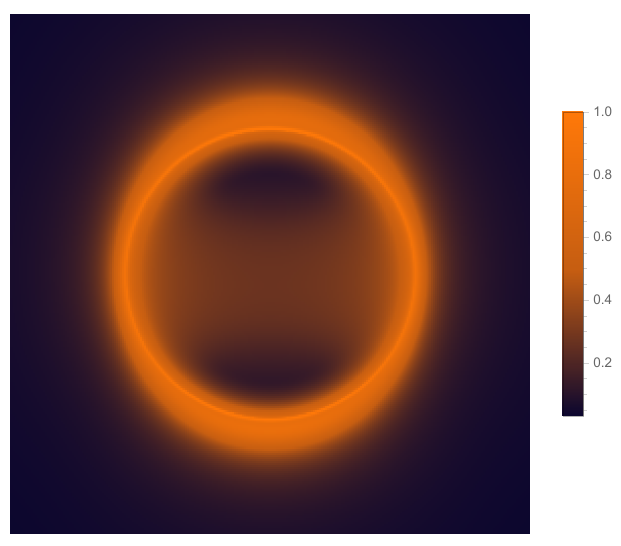}}

\subfigure[$b^r/b^\phi=0.5\,,\quad b^\theta=0$]{\includegraphics[scale=0.45]{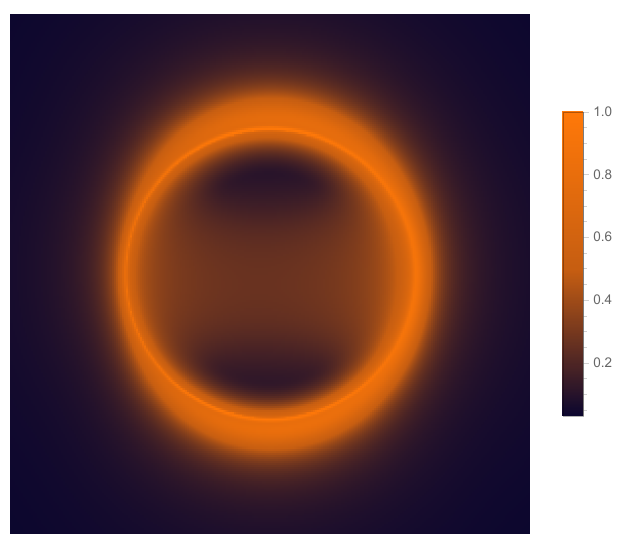}}
	\subfigure[$b^\theta/b^\phi=0.5\,,\quad b^r=0$]{\includegraphics[scale=0.45]{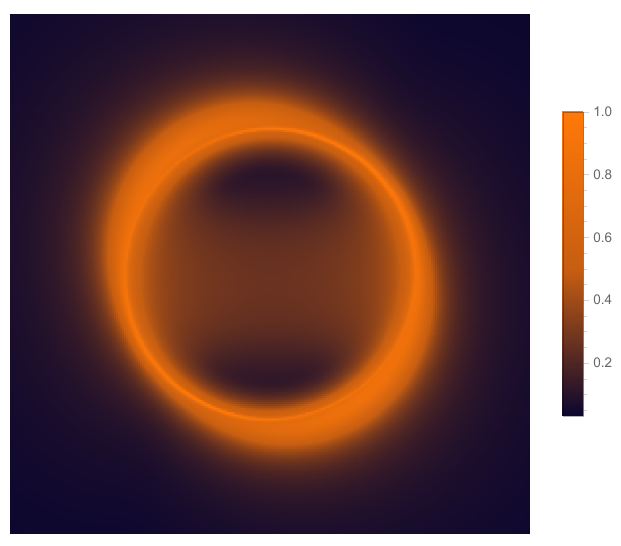}}
	\subfigure[$b^r/b^\phi=b^\theta/b^\phi=0.5$]{\includegraphics[scale=0.45]{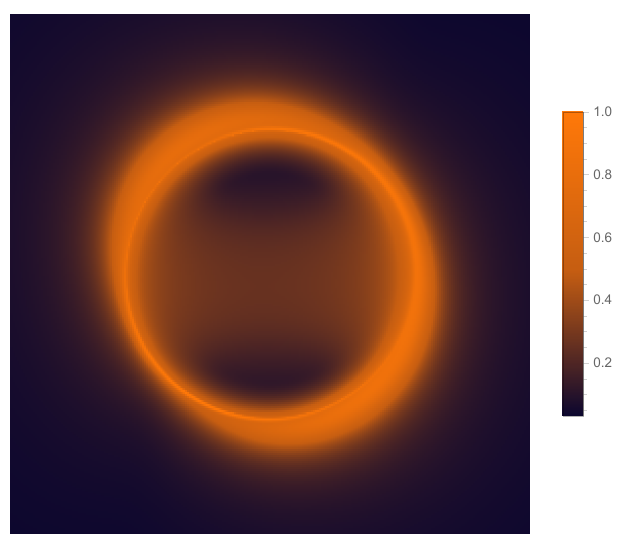}}
    
    \caption{Intensity maps of the KZ black hole in the RIAF model with anisotropic emission for different magnetic field configurations. The accretion flow follows the infalling motion, the observing frequency is 230 GHz.}
    \label{fig:15}
\end{figure}

\bibliographystyle{utphys}
\bibliography{reference}

\end{document}